%% file: main_oja.tex
\documentclass[apj,twocolumn,twocolappendix,numberedappendix,appendixfloats]{openjournal}
\usepackage{float}
\usepackage{graphicx}
\usepackage{amsmath}
\usepackage{url}
\usepackage[breaklinks,colorlinks,citecolor=blue,linkcolor=blue,urlcolor=blue]{hyperref}
\usepackage{listings}
\providecommand{\doi}[1]{doi: \href{https://doi.org/#1}{\nolinkurl{#1}}}
\usepackage{lineno}
 
\input{journals}

\begin{document}

\title{A heterogeneous set of unusual quasars selected from astrometry and broad-band photometry}
\shortauthors{Cheng et al.}

\author{
C.~Cheng$^{1,2}$,
K.~Valeckas$^{1,2,3,4}$,
L.~Christensen$^{1,2}$,
A.~Henderson de la Fuente$^{3,5}$,
L.~Fuglsang$^{2,3,6}$,
V.~Vuolteenaho$^{3,7}$,
N.~Pyykkinen$^{3,8}$,
R.~H. Rasmussen$^{3,9}$,
S.~Bijavara Seshashayana$^{3,10}$,
J.~P.~U.~Fynbo$^{1,2}$\footnotemark[*]
}

\affiliation{$^{1}$Cosmic DAWN Center}
\affiliation{$^{2}$Niels Bohr Institute, Copenhagen University, Jagtvej 155, DK-2200 Copenhagen N, Denmark}
\affiliation{$^{3}$Nordic Optical Telescope, Rambla Jos{\'e} Ana Fern{\'a}ndez P{\'e}rez 7, ES-38711 Br{\~e}na Baja, Spain}
\affiliation{$^{4}$European Southern Observatory, Alonso de Córdova 3107, Vitacura, Región Metropolitana, Chile}
\affiliation{$^{5}$Department of Physics, University of Oxford, Keble Road, Oxford, OX1 3RH, UK} 
\affiliation{$^{6}$Centre de Recherche Astrophysique de Lyon, Université Lyon 1, 9 Avenue Charles André, 69561 Saint Genis Laval, France}
\affiliation{$^{7}$Space Physics and Astronomy Research Unit, University of Oulu, P.O. Box 3000, FI-90014 University of Oulu, Finland}
\affiliation{$^{8}$Department of Physics and Astronomy, University of Turku, FI20014 Turku, Finland}
\affiliation{$^{9}$Department of Physics and Astronomy, Aarhus University, Ny Munkegade 120, DK-8000, Aarhus C, Denmark}
\affiliation{$^{10}$Materials Science and Applied Mathematics, Malmö University, SE-205 06 Malmö, Sweden}

\footnotetext[*]{Corresponding author. jfynbo@nbi.ku.dk}
 
\begin{abstract}
The aim of this study is to explore methods for how to identify new quasars with different properties with regard to reddening, redshift, the presence of broad absorption lines from outflows, or other rare properties affecting the broad-band colours. We use Gaia-based astrometric pre-selection to study only sources that have zero parallax and zero proper motion within errors. We subsequently match these sources with information from wide-area photometric surveys and build colour-colour diagrams. Based on these diagrams we select targets for follow-up spectroscopy. In the subsequent analysis we also apply observations from SPHEREx to firmly establish redshifts. 
We secure spectroscopic observations for 24 sources. Of these, three are found to be stars and the remaining 21 are classified as quasars spanning a redshift range from $z=0.139$ to $z=3.732$. The majority of the sources located away from the normal quasar locus in colour-colour diagrams are found to be broad-absorption-line quasars. A single source, the lowest redshift source at $z=0.139$, is reminiscent of the so-called Little Red Dots found at high redshifts with JWST. Using the simple procedure followed here it is possible to locate previously undiscovered quasars with interesting properties, in particular Broad Absorption Line quasars and quasars with unusual extinction.
\end{abstract}

%%%%%%%%%%%%%%%%%%%%%%%%%%%%%%%%%%%%%%%%%%%%%%%%%%%%%%%%%%%%%%
\section{Introduction}
Since their discovery in the 1960s, quasars have played a central role in extragalactic astronomy, serving as powerful probes of the early Universe and as tracers of accretion onto supermassive black holes at the centres of galaxies \citep{fan2006}. Powered by accretion onto supermassive black holes \citep{rees1984}, quasars provide key insights into the physics of active galactic nuclei (AGN), the growth of supermassive black holes, and their interplay with galaxy evolution, as well as the properties of the intervening intergalactic medium \citep{rees1984,hopkins2006,meiksin2009}.

Historically, quasars have been identified using a variety of selection techniques. Early surveys primarily relied on radio emission, leading to the discovery of the first quasar samples \citep{1963Natur.197.1040S}. While radio selection is largely insensitive to dust extinction, it is intrinsically biased towards the radio-loud population, which represents only a small fraction of the overall quasar population \citep{1965ApJ...141.1560S}. Hence it was realized that radio-selection alone cannot provide a complete census of quasars.

Subsequently, colour-based selection methods became the dominant approach, particularly with the advent of wide-area optical surveys. By exploiting the distinctive colours of quasars relative to stars, optical selection has enabled the construction of large quasar catalogs with up to a million sources. Extensions to the near- and mid-infrared have further improved selection efficiency and helped recover quasars missed in optical surveys. However, all colour-based techniques ultimately depend on the observed spectral energy distributions of sources and therefore remain susceptible to colour-dependent biases, especially for quasars affected by dust reddening or strong absorption features such as Broad Absorption Lines \citep[BALs,][]{2009ApJ...692..758G}.

An alternative and complementary approach to quasar selection is provided by astrometric methods \citep[e.g.,][]{1986PASP...98..285K, 2002A&A...392..851M}. This approach exploits the negligible parallax and proper motion of quasars, e.g. using very precise observations from the Gaia mission \citep{2016A&A...595A...1G}, to identify candidates independently of their spectral energy distributions or colours. Previous studies have demonstrated the feasibility of this approach using both simulated and real Gaia data, confirming that astrometric selection can recover quasars that are underrepresented in traditional colour-based samples \citep{2015A&A...578A..91H,2018A&A...615L...8H,2020A&A...644A..17H,2019A&A...625L...9G,2020A&A...634A.111F}.

In this work, we use a criterion of zero proper motion and a parallax of zero as an astrometric pre-selection and sample sources from different regions of colour–colour space to investigate the connection between photometric properties and spectroscopic characteristics. This selection is not optimized for completeness or sample purity, but is instead designed to explore the diversity of spectroscopic characteristics among astrometrically selected candidates. 

The paper is organized in the following way: In Sect.~\ref{selection} we describe how the targets of this study have been selected. In Sect.~\ref{observations} we describe the observations, which mostly come from a summer school in August 2024, with additional observations extending the sample in 2025. In Sect.~\ref{results} we describe our results, which mainly consists of analysis of the spectra and classification of the sources in terms of object class (star or quasar, which type of quasar),
extinction and redshift. In Sect.~\ref{discussion} we discuss the implications of our work and in Sect.~\ref{conclusions} we provide our conclusions.
%%%%%%%%%%%%%%%%%%%%%%%%%%%%%%%%%%%%%%%%%%%%%%%%%%%%%%%%%%%%%%

\section{Selection Strategy}
\label{selection}

The target selection in this work was designed to investigate the spectroscopic properties of quasar candidates occupying different regions of colour–colour space. Without the inclusion of astrometric information such a study is nearly impossible as some regions of colour-space are swamped by stellar sources making it very difficult to pick out the rare non-stellar sources in these regions of colr-space. We have adopted a selection strategy that combines an astrometric pre-selection, and hence rejection of the large majority of stellar sources, with targeted sampling based on photometric properties.

\subsection{Astrometric pre-selection}
We first selected sources from the Gaia catalogue with proper motions and parallaxes consistent with zero within the measurement uncertainties. Such sources are expected to be dominated by extragalactic objects and therefore constitute a natural pool of quasar candidates. 
Our SQL-query for the astrometric preselection is provided in Fig.~\ref{selectioncriteria}. This set of objects we will refer
to as our master catalogue.

\begin{figure}
\centering
\begin{minipage}{0.9\linewidth}
\begin{lstlisting}
SELECT TOP 50000000
    source_id, ra, dec, l, b,
    pmra, pmra_error,
    pmdec, pmdec_error,
    parallax, parallax_error,
    phot_g_mean_mag,
    phot_g_mean_flux,
    phot_g_mean_flux_error,
    phot_bp_mean_mag,
    phot_rp_mean_mag,
    bp_rp,
    ref_epoch
FROM gaiaedr3.gaia_source AS gaia
WHERE
    gaia.ra >= 270
    AND gaia.ra <= 359
    AND gaia.dec > 0
    AND gaia.phot_g_mean_mag <= 19
    AND ABS(gaia.parallax/gaia.parallax_error) <= 2
    AND (
        POWER(gaia.pmra/gaia.pmra_error,2)
      + POWER(gaia.pmdec/gaia.pmdec_error,2)
    ) <= 4
\end{lstlisting}
\end{minipage}
\caption{The SQL query we have used to pre-select stationary and distant Gaia EDR3 sources for this mini-survey.}
\label{selectioncriteria}
\end{figure}

\subsection{Further down-selection based on position in colour–colour space}
\label{colour–colour selection}
These astrometrically selected sources were then cross-matched with photometric data from the Sloan Digital Sky Survey \citep[SDSS, bands $u, g, r, i, z$,][]{SDSS}, as well as complementary near- and mid-infrared data from the Two Micron All Sky Survey \citep[2MASS, bands $J, H, K$,][]{2MASS} and the Wide-field Infrared Survey Explorer \citep[WISE, bands $W1, W2, W3, W4$,][]{WISE}. 

\begin{figure*}[htbp]
\centering
\includegraphics[width=0.48\textwidth]{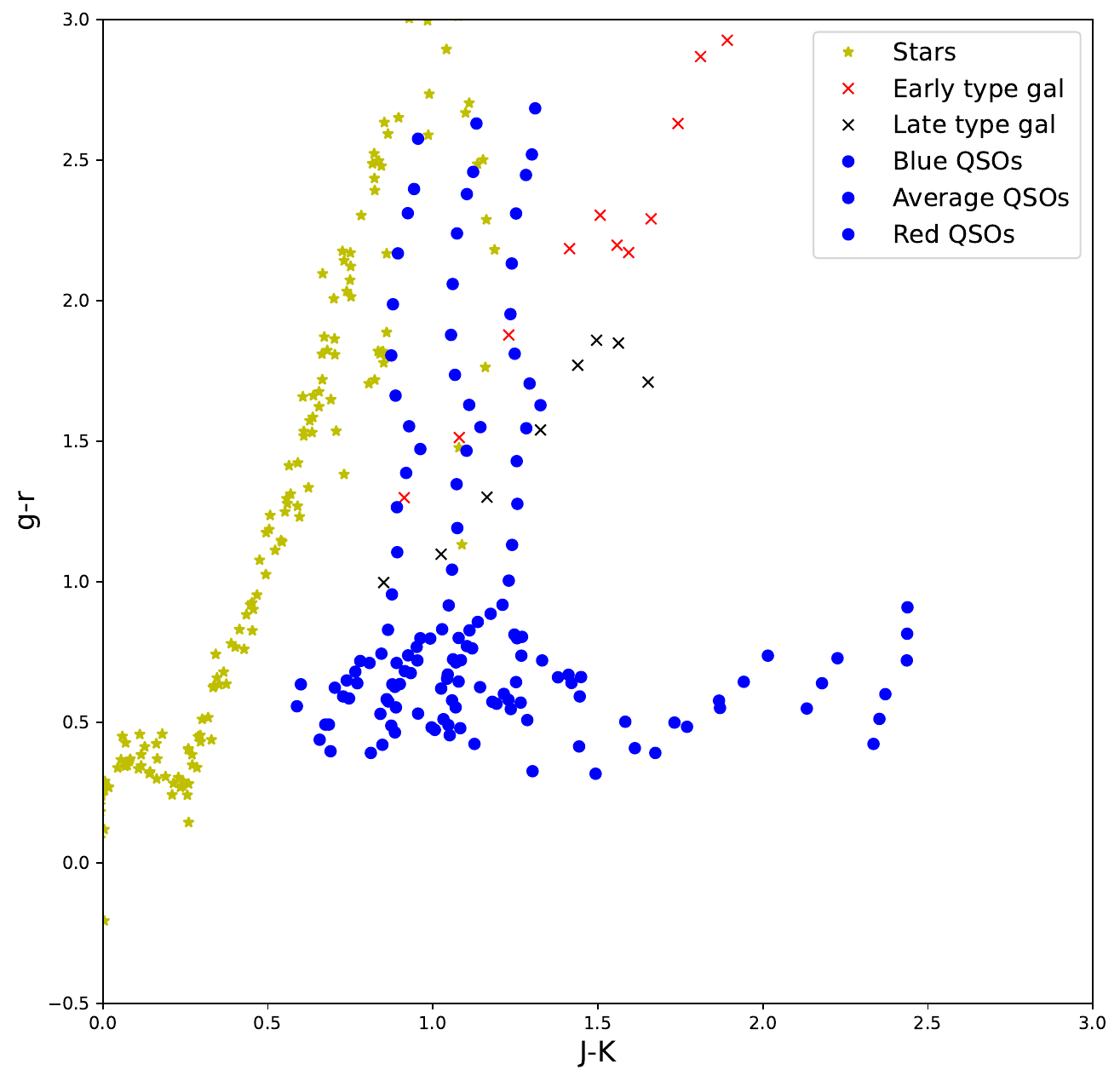}
\hfill
\includegraphics[width=0.48\textwidth]{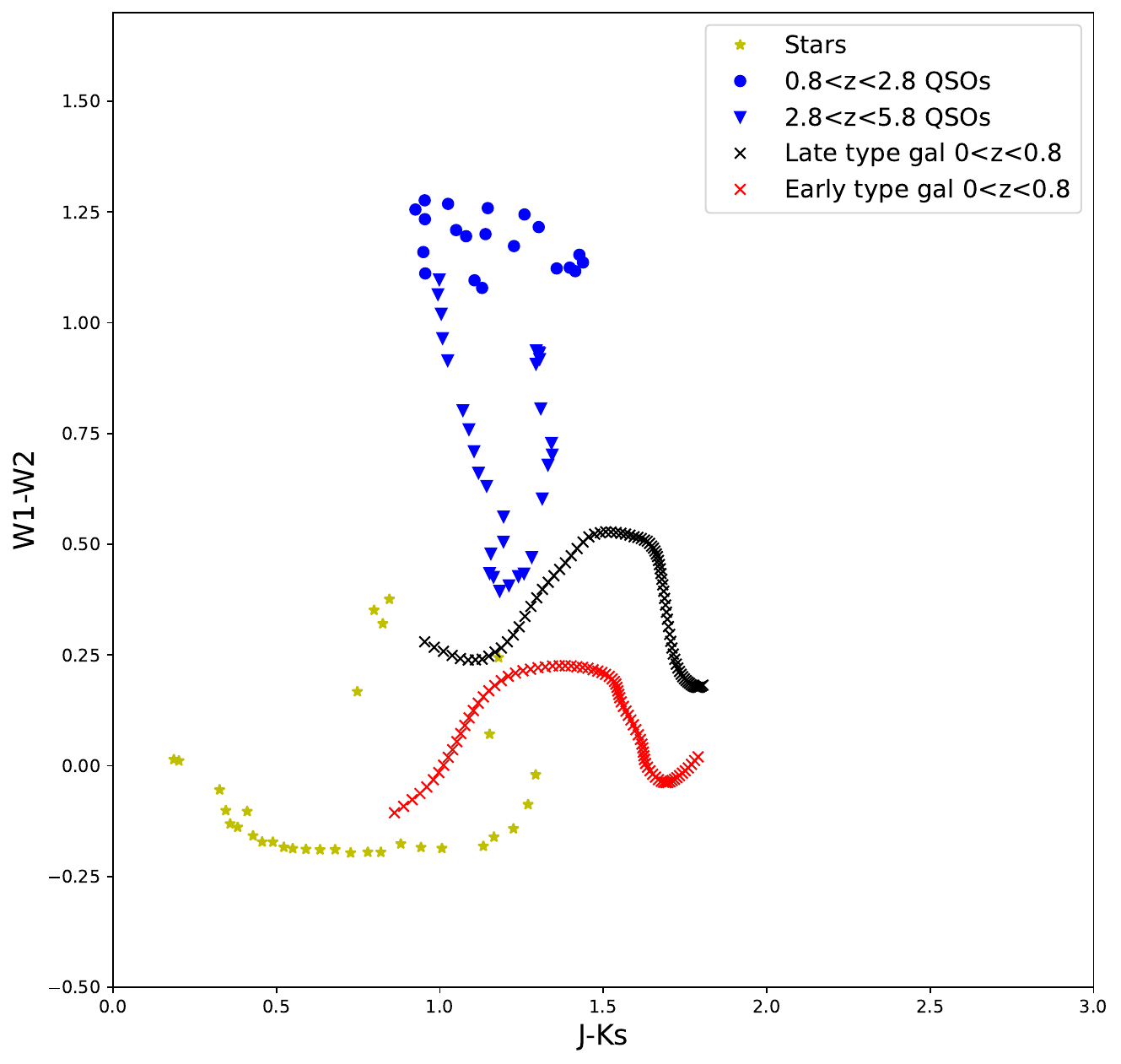}
\vspace{0.5em}

\includegraphics[width=0.48\textwidth]{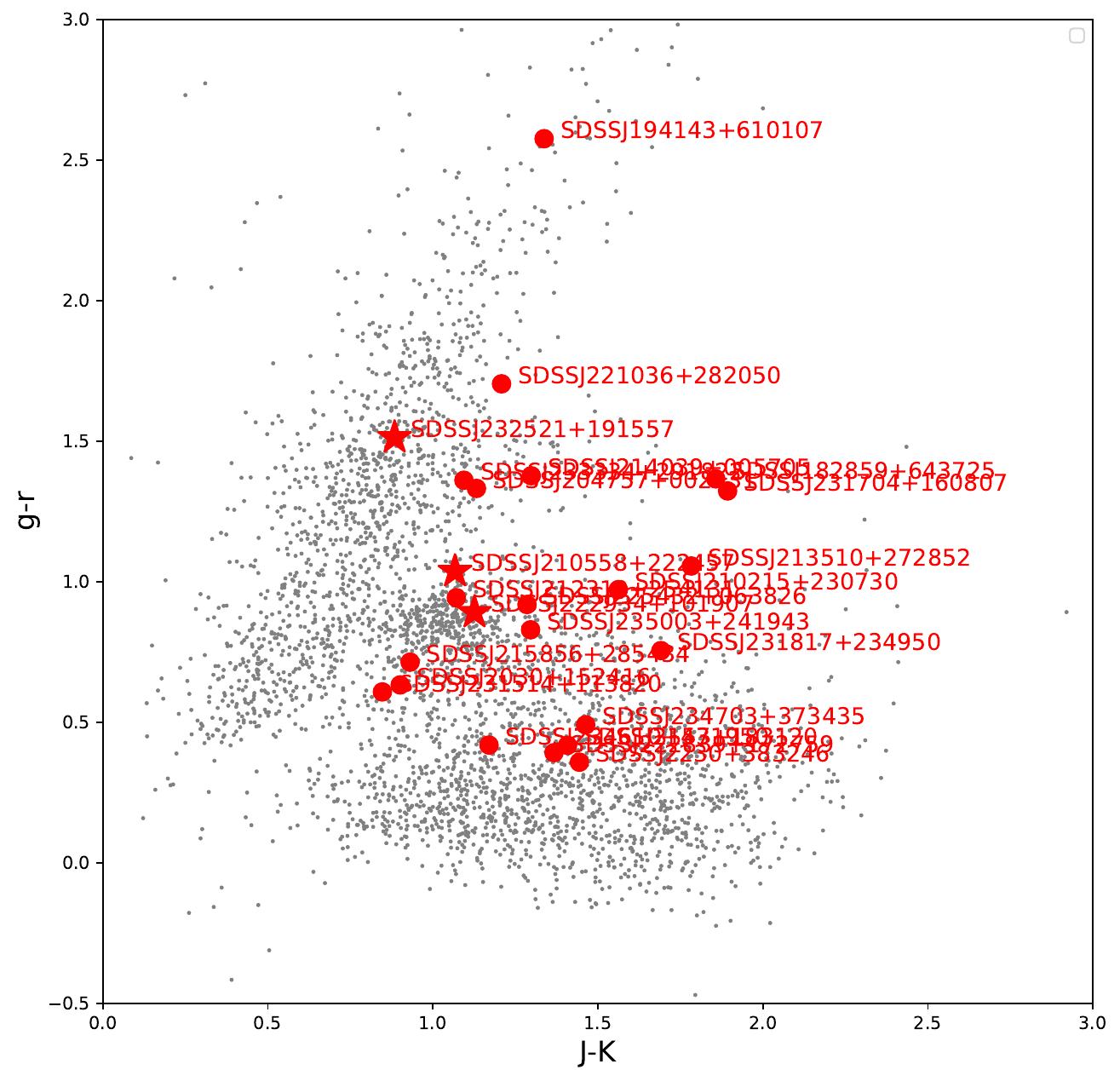}
\hfill
\includegraphics[width=0.48\textwidth]{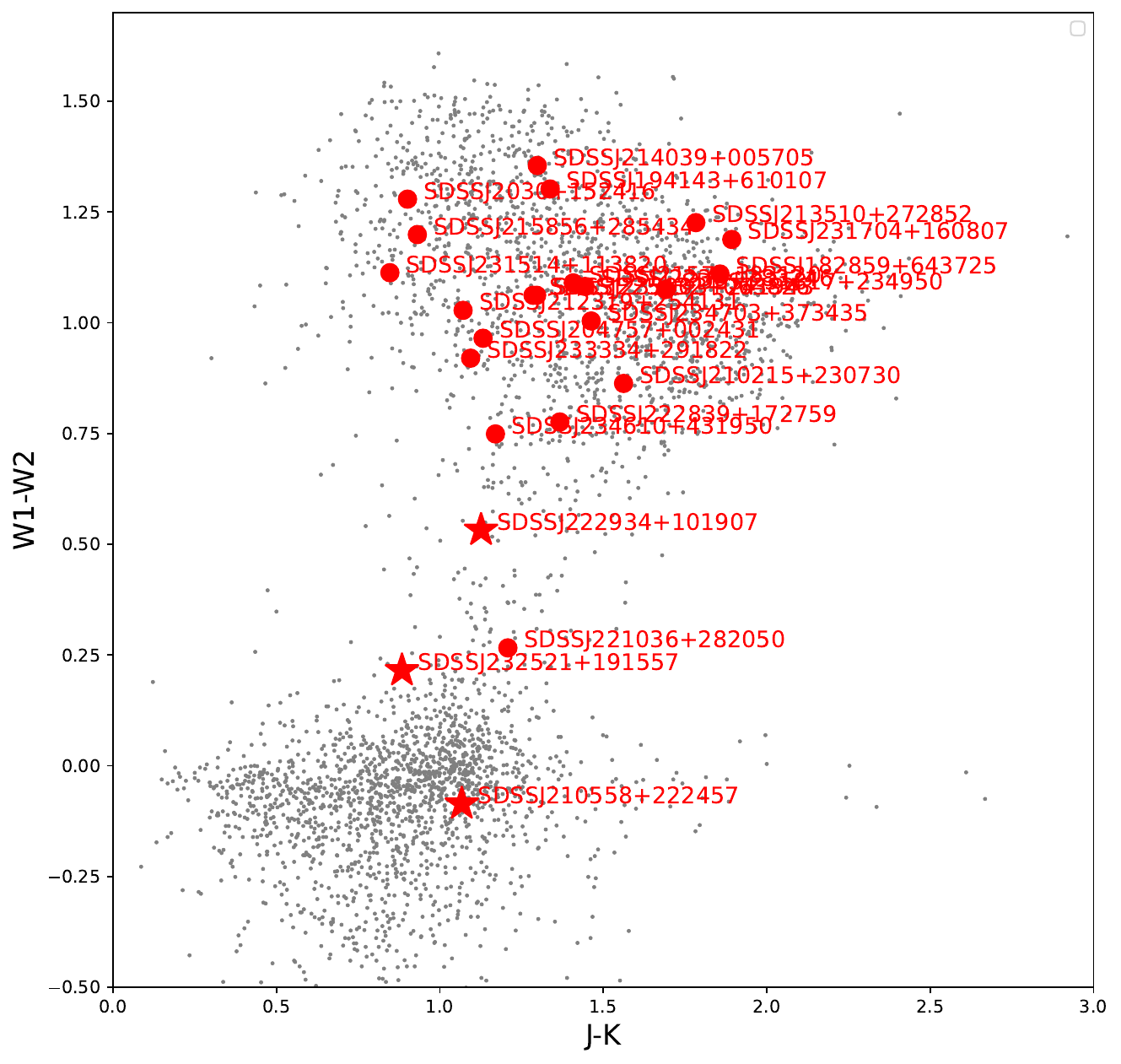}
\caption{Colour--colour diagrams. 
{\it Top panels:} Theoretical colour-tracks for stars, galaxies and quasars.
{\it Top Left:} Colours derived using the templates from \citet{2006MNRAS.367..454H} as described in \citet{2017A&A...606A..13F}. {\it Top Right:} For this diagram we calculate synthetic colours using templates from \citet{2010ApJ...713..970A,2013A&A...553A...6H,2016A&A...585A..87S}.
{\it Bottom panels:} colours of objects in the master catalogue. Confirmed quasars are shown with filled circles and stars with an asterisk.}
\label{fig:colour}
\end{figure*}

Using the available photometry, we constructed colour–colour diagrams for further down-selection of targets for spectroscopic follow-up.
We use two main colour-colour diagrams, namely $g-r$ versus $J-K$
and $W1-W2$ vs. $J-K$. These diagrams provide very useful separations of stars, galaxies and quasars \citep[e.g.,][]{2017A&A...606A..13F,
2019A&A...625L...9G}. In colour--colour space, the expected distributions of stars and quasars can be understood from theoretical templates. As illustrated in Figure \ref{fig:colour}, stars follow a narrow and well-defined track in the $g-r$ versus $J-K$ diagram, reflecting the relatively uniform spectral energy distributions of stellar populations. In contrast, quasars do not occupy a single linear sequence but instead populate a broader and more structured locus with a finite width due to redshift, which moves the strong emission lines in and out of the various filters and removes light due to Lyman-$\alpha$ 
forest blanketing at $z\gtrsim2.5$, and different levels of extinction. In particular, the position of the H$\alpha$ line to a large extent explains the spread in $J-K$ colour for quasars.  

%\begin{figure}[htbp]
%    \centering
%    \includegraphics[width=0.45\textwidth]{Figures/grJK.pdf}
%    \caption{Theoretical colour-tracks for stars, galaxies and quasars derived using the templates from %\citet{2006MNRAS.367..454H} as described in \citet{2017A&A...606A..13F}. For the WISE bands we calculate 
%    synthetic colours using templates from \citet{2010ApJ...713..970A,2013A&A...553A...6H,2016A&A...585A..87S,}.}
%    \label{fig:theory}
%\end{figure}

Although the stellar and quasar loci are largely separated, Fig.~\ref{fig:colour} also shows that partial overlap is expected in specific regions of colour--colour space. On the left side of the $g-r$ vs. $J-K$ diagram, at relatively blue near-infrared colours $(J-K \lesssim 0.6)$, the stellar locus forms a comparatively clean and “safe” region dominated by stars. At larger $J-K$ values, stellar templates continue to appear but increasingly overlap with the quasar locus, particularly with blue quasar tracks. For quasars, a relatively safe region is found in the lower part of the quasar locus, where little overlap with the stellar sequence is expected. In the $W1-W2$ vs. $J-K$ diagram there is also overlap between different classes of objects 
at $W1-W2 \approx 0.5$. These theoretical predictions indicate that although much of colour–colour space allows fairly robust source classification, certain transition regions are more ambiguous.

Targets were selected from different regions of these diagrams for spectroscopic follow-up, including both regions where stars and quasars are well separated and regions where their distributions overlap.

Rather than optimizing the selection for sample purity or completeness, this approach was adopted to probe the diversity of spectroscopic properties associated with different locations in colour–colour space. By sampling both photometrically “clean” regions and more ambiguous transitional regions, this strategy allows a direct assessment of the extent to which photometric location reflects the properties such as redshift, extinction and the presence of BAL features of the sources.

\begin{figure}[th]
\centering
\includegraphics[scale=0.60]{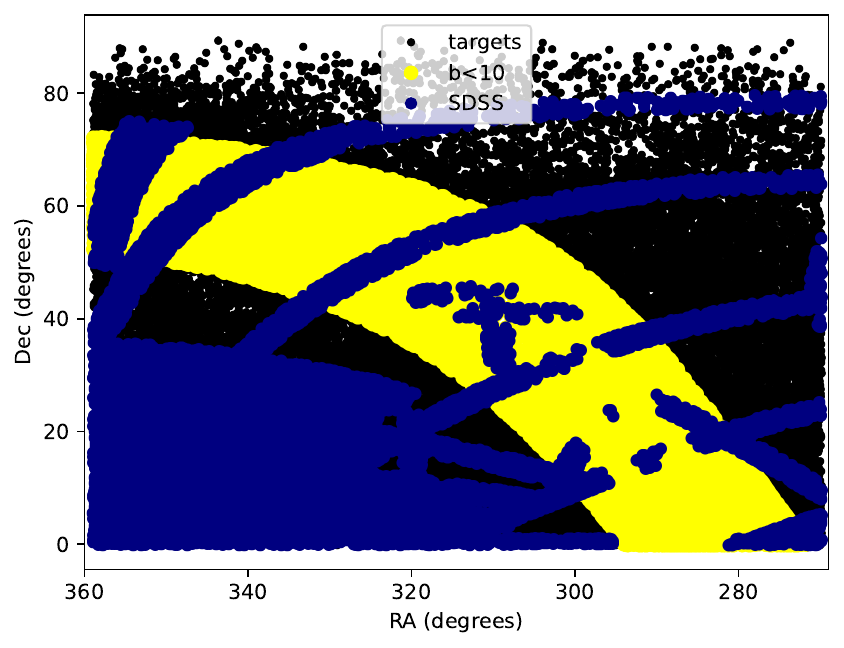}
\caption{Our targets are selected to be observable from La Palma in August. Targets are selected to be above a Galactic
height of 10 degrees and from the parts of the northern summer sky covered by SDSS. For these regions there is also 
available photometry from 2MASS and WISE.
}
\label{fig:ukidss}
\end{figure}

\section{Observations}
\label{observations}

All spectroscopic observations presented in this work were made with the 2.56-m NOT located on the Canarian island of La Palma, using the Andalucia Faint Object Spectrograph and Camera (ALFOSC). ALFOSC was chosen for its efficiency in observing faint sources, making it well suited for low- to intermediate-resolution long-slit spectroscopy of quasar candidates.

Most targets were observed using grism 4, which provides a wavelength coverage of approximately 3200–9600 {\AA} with a spectral resolution of slightly below 400 when using a 1" slit. This setup is well suited for identifying prominent emission and absorption features in quasar spectra over a wide redshift range. A subset of targets was re-observed using grism 20, covering the wavelength range 5650–10150 Å, to obtain follow-up spectra with improved sensitivity at longer wavelengths and approximately double the resolution compared to grism 4. Both grism 4 and grism 20 suffer from about 5\% 2nd-order contamination redward of 9000 \AA. This combination of grisms offers broad wavelength coverage while maintaining a suitable balance between spectral resolution and signal-to-noise ratio for reliable spectral classification and redshift determination. A complete list of the observed targets, together with their coordinates and observational parameters, is provided in Table \ref{tab:exposure_targets} and Table \ref{tab:grism20_targets}. 

The spectroscopic data were reduced using standard procedures for long-slit spectroscopy, including bias subtraction, flat-field correction, wavelength calibration, sky subtraction, spectral extraction, and flux calibration. Data reduction and analysis were performed using Python-based routines as implemented in the Python-package PyLongslit \citep{Pylongslit}, which is available on Github\footnote{\tiny{\url{https://kostasvaleckas.github.io/PyLongslit/index.html}}}. 

\begin{table*}[htbp]
\small
\renewcommand{\arraystretch}{0.9}
\centering
\caption{All targets observed with Grism 4.}
\label{tab:exposure_targets}
\begin{tabular}{lccccc}
\hline\hline
Target & RA (J2000) & Dec (J2000) & Date &  Exposure Time & Airmass \\
                & \textit{[hr:min:sec]} & \textit{[deg:min:sec]} &   & \textit{[s]} & \\
\hline
SDSS\,J182859+643725 & 18:28:59.34 & +64:37:25.30 & 2024-08-10 & 2$\times$300 & 1.34\\
SDSS\,J194143+610107 & 19:41:43.66 & +61:01:07.80 & 2024-08-05 & 2$\times$300 & 1.22 \\
SDSS\,J203008+152416 & 20:30:08.82 & +15:24:16.00 & 2024-08-05 & 2$\times$300 & 1.30 \\
SDSS\,J204757+002431 & 20:47:57.18 & +00:24:31.50 & 2024-08-10 & 2$\times$500 & 1.14 \\
SDSS\,J210215+230730 & 21:02:15.48 & +23:07:30.72 & 2025-07-31 & 2$\times$600 & 1.01 \\
SDSS\,J210558+222457 & 21:05:58.53 & +22:24:57.00 & 2024-08-07 & 2$\times$300 & 1.06 \\
SDSS\,J212319+254131 & 21:23:19.53 & +25:41:31.20 & 2024-08-10 & 2$\times$300 & 1.02 \\
SDSS\,J213510+272852 & 21:35:10.10 & +27:28:52.20 & 2024-08-09 & 2$\times$400 & 1.21 \\
SDSS\,J214039+005705 & 21:40:39.85 & +00:57:05.80 & 2024-08-10 & 2$\times$300 & 1.13 \\
SDSS\,J215731+183120 & 21:57:31.94 & +18:31:20.40 & 2024-08-09 & 2$\times$200 & 1.02 \\
SDSS\,J215856+285434 & 21:58:56.42 & +28:54:34.50 & 2024-08-06 & 2$\times$300 & 1.00 \\
SDSS\,J221036+282050 & 22:10:36.32 & +28:20:50.65 & 2025-08-11 & 2$\times$1200 & 1.06\\ 
SDSS\,J222839+172759 & 22:28:39.76 & +17:27:59.57 & 2025-07-18 & 2$\times$600 & 1.02 \\
SDSS\,J222943+101907 & 22:29:34.34 & +10:19:07.99 & 2024-08-09 & 2$\times$300 & 1.20 \\
SDSS\,J223052+383246 & 22:30:22.53 & +38:32:46.60 & 2024-08-09 & 2$\times$300 & 1.09 \\
%SDSS\,J224129+024807 & 22:41:29.03 & +02:48:07.23 &  2025-07-26 & 2$\times$600 \\
SDSS\,J225432+063826 & 22:54:32.27 & +06:38:26.04  & 2025-08-22 & 2$\times$900 & 1.08\\
SDSS\,J231514+113820 & 23:15:14.30 & +11:38:20.20 & 2024-08-24 & 2$\times$600 & 1.05 \\
%SDSS\,J223511+1138 & 22:35:11.34 & +11:38:00.90 & 300 \\
SDSS\,J231704+160807 & 23:17:04.31 & +16:08:07.30 & 2024-08-09 & 2$\times$300 & 1.14 \\
SDSS\,J231817+234950 & 23:18:17.92 & +23:49:50.10  & 2025-08-22 & 2$\times$1200 & 1.20\\
SDSS\,J232521+191557 & 23:25:21.01 & +19:15:57.06 & 2024-08-10 & 2$\times$300 & 1.10 \\
SDSS\,J233334+291822 & 23:33:34.32 & +29:18:22.08  & 2025-09-20 & 2$\times$1400 & 1.29 \\
SDSS\,J234610+431950 & 23:46:10.77 & +43:19:50.61 & 2024-08-09 & 2$\times$300 & 1.06 \\
SDSS\,J234703+373435 & 23:47:03.04 & +37:34:35.10 & 2024-08-25 & 2$\times$900 & 1.15 \\
SDSS\,J235003+241943 & 23:50:03.81 & +24:19:43.79  & 2025-07-30 & 2$\times$900 & 1.02\\
\hline
\end{tabular}
\end{table*}

\begin{table*}[htbp]
\small
\renewcommand{\arraystretch}{0.9}
\centering
\caption{All targets observed with Grism 20.}
\label{tab:grism20_targets}
\begin{tabular}{lccccc}
\hline\hline
Target & RA (J2000) & Dec (J2000) & Date & Exposure Time & Airmass \\
                & \textit{[hr:min:sec]} & \textit{[deg:min:sec]} & & \textit{[s]} & \\
\hline
SDSS\,J231704+160820 & 23:17:04.31 & +16:08:20.70 & 2024-08-25 & 2$\times$600 & 1.02 \\
SDSS\,J182859+643725 & 18:28:59.34 & +64:37:25.30 & 2024-08-10 & 2$\times$500 & 1.32 \\
SDSS\,J204757+002034 & 20:47:57.18 & +00:20:34.50 & 2024-09-05 & 2$\times$600 & 1.15 \\
\hline
\end{tabular}
\end{table*}

%%%%%%%%%%%%%%%%%%%%%%%%%%%%%%%%%%%%%%%%%%%%%%%%%%%%%%%%%%%%%%
\section{Results}
\label{results}

\subsection{Spectroscopic classification}

Spectroscopic observations were obtained for 24 candidates. Based on the inspection of the one-dimensional spectra and comparison with quasar templates, 21 objects are confirmed as quasars, the remaining 3 targets are classified as stellar contaminants. To classify the objects we plot the optical spectra from
our own observations together with photometry from SDSS; UKIDSS, and WISE and with spectroscopic data from SPHEREx \citep{2026ApJ...999..139B}. The SPHEREx spectra are acquired using the SPHEREx pipeline available on Github\footnote{\tiny{\url{https://github.com/andrijazupic/SPHEREx\_pipeline/tree/main}}}. WISE photometry is in particular interesting as the
The criterion $W1 - W2 > 0.8$ is a robust selector for quasars at $z<2.8$ \citep{2012ApJ...753...30S}. At higher redshifts $W1 - W2$ drops and blends with the $W1-W2$ colours of stars and galaxies (see Fig.~\ref{fig:colour}).

For quasar candidates, redshifts and dust extinction values were estimated by comparing the observed spectra with a representative quasar template, for which we adopt the \citet{2016A&A...585A..87S} quasar template, reddened using an SMC-like extinction curve as quantified by \citet{1992ApJ...395..130P}. For BAL quasars, where emission lines may be strongly suppressed, redshift determinations primarily rely on absorption features and on emission lines in the SPHEREx spectral region (H$\alpha$ in particular). 

We also cross-check with the FIRST and LOFAR survey for emission at radio wavelengths \citep{2015ApJ...801...26H,2023A&A...673A.165D}. 

The classifications of all targets, together with the estimated redshifts and extinction values, are listed in Table~\ref{tab:quasar_sample_basic}.  Fig.~\ref{fig:colour} shows the positions of all targets in the $g-r$ versus $J-K$ diagram, with the spectroscopic classifications indicated by different symbols and the redshifts represented by colour.

\begin{table}[htbp]
\centering
\caption{Quasar sample: classification, redshift and extinction.}
\label{tab:quasar_sample_basic}
\begin{tabular}{l l l l}
\hline\hline
    Target & Type & Redshift & A$_\mathrm{V}$ \\
\hline
SDSS\,J182859+643725 & peculiar/BAL      & 0.609 & 1.4 \\
SDSS\,J194143+610107 & BAL      & 1.042 & 1.4 \\
SDSS\,J203008+152416 & BAL      & 1.741 & 0.0 \\
SDSS\,J204757+002431 & BAL      & 1.85   & 0.80 \\
SDSS\,J210215+230730 & normal   & 0.139  & ? \\
SDSS\,J210558+222457 & star     & 0.0    & 0.0 \\
SDSS\,J212319+254131 & BAL      & 2.520   & 0.1 \\
SDSS\,J213510+272852 & BAL      & 2.43 & 0.20 \\
SDSS\,J214039+005705 & peculiar & 1.1555 & 0.55 \\
SDSS\,J215731+183120 & normal   & 2.02   & 0.15 \\
SDSS\,J215856+285434 & BAL      & 1.553  & 0.0 \\
SDSS\,J221036+282050 & normal   & 3.732  & 0.2 \\
SDSS\,J222839+172759 & BAL  & 2.932  & 0.25 \\
SDSS\,J222934+101907 & star     & 0.0    & 0.0 \\
SDSS\,J223030+383246 & normal  & 2.26   & 0.0 \\
SDSS\,J225432+063826 & peculiar & 2.698 & 0.40 \\
SDSS\,J231514+113820 & BAL     & 2.895   & 0.2 \\
SDSS\,J231704+160807 & peculiar  & 0.464  & 1.7 \\
SDSS\,J231817+234950 & normal  & 0.570 & 1.50 \\
SDSS\,J232521+191557 & star    & 0.0    & 0.0 \\
SDSS\,J233334+291822 & BAL    & 1.675  & 0.60 \\
SDSS\,J234610+431950 & normal   & 3.26   & 0.0 \\
SDSS\,J234703+373435 & normal   & 3.03   & 0.20 \\
%SDSS\,J224129+024807 &   &   & \\
SDSS\,J235003+241943 & peculiar  & 0.692  & 0.3 \\
\hline
\end{tabular}
\end{table}

\subsection{Notes on individual objects}

\input{notes_individual_objects}

\subsection{The confirmed quasars}
\begin{figure}[htbp]
    \centering
    \includegraphics[width=0.9\columnwidth]{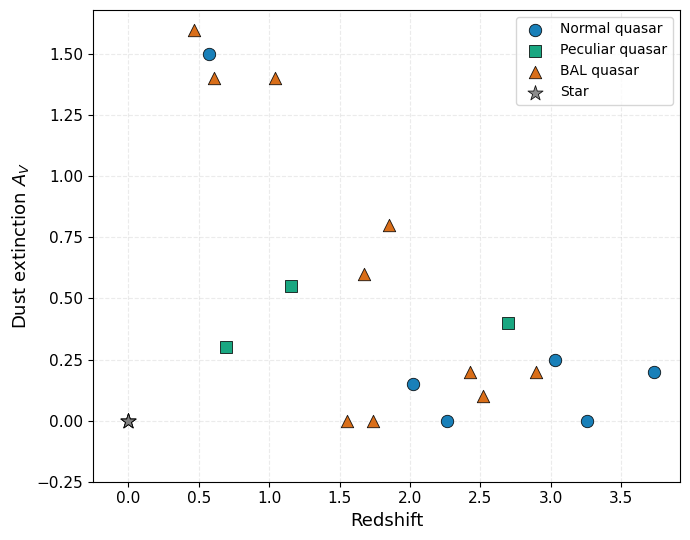}
    \caption{Estimated dust extinction, $A_V$, as a function of 
    redshift for the confirmed quasars. Different marker shapes indicate 
    the spectral classifications of the quasars. The most highly 
    extinguished objects are found predominantly at relatively low 
    redshifts.}
    \label{fig:Av_redshift}
\end{figure}

The 21 confirmed quasars span a wide redshift range from  $z=0.139$ to $z=3.732$. The estimated dust extinction values cover a range of $A_V = 0$ to $A_V = 1.7$. As expected, the most dust obscured objects are at the lowest redshifts as shown in 
Fig.~\ref{fig:Av_redshift}. This is what we expect due to Malmquist bias and due to the fact that it requires higher extinction to redden a spectrum in the observed optical range for a lower redshift source compared to a higher redshift source. 

Based on their spectral characteristics, the confirmed quasars are classified into 7 normal quasars (including some that are obscured) and 11 broad absorption line (BAL) quasars, and three peculiar 
quasars with very weak emission lines. Among the BAL quasars, four objects are further classified as FeLoBAL quasars, identified by the presence of strong \ion{Fe}{2} absorption features in their spectra.

Fig.\ref{fig:colour} shows the distribution of all observed candidates in the $g-r$ versus $J-K$ colour-colour diagram. 
The quasars occupy a broad region of the colour-colour space, extending beyond the main locus of optically selected quasars. BAL quasars are predominantly located at higher $g-r$ values, occupying the upper part of the diagram. In contrast, the normal quasars cluster toward lower $g-r$ values, close to the region typically populated by optically selected quasars.

\subsection{Stars}
Among the 17 spectroscopically observed candidates, three objects are classified as late-type stars (stellar class K and M).
These stars are located close to the boundary between the quasar and stellar loci in the $g-r$ versus $J-K$ colour--colour diagram, consistent with their spectroscopic identification.

%%%%%%%%%%%%%%%%%%%%%%%%%%%%%%%%%%%%%%%%%%%%%%%%%%%%%%%%%%%%%%
\section{Discussion}
\label{discussion}

\subsection{Theoretical and empirical structure of the colour–colour diagram}

As discussed in Sect.~\ref{colour–colour selection}, the theoretical colour tracks in the $g-r$ vs. $J-K$ diagram derived from spectral templates predict a relatively narrow stellar locus, a lower quasar locus, and intermediate regions in which the two populations approach or overlap. These predictions formed the basis for the division of sources into relatively clean stellar and quasar regions and more ambiguous regions with significant overlap of
sources of different nature.

\begin{figure}[htbp]
\centering
\includegraphics[width=0.45\textwidth]{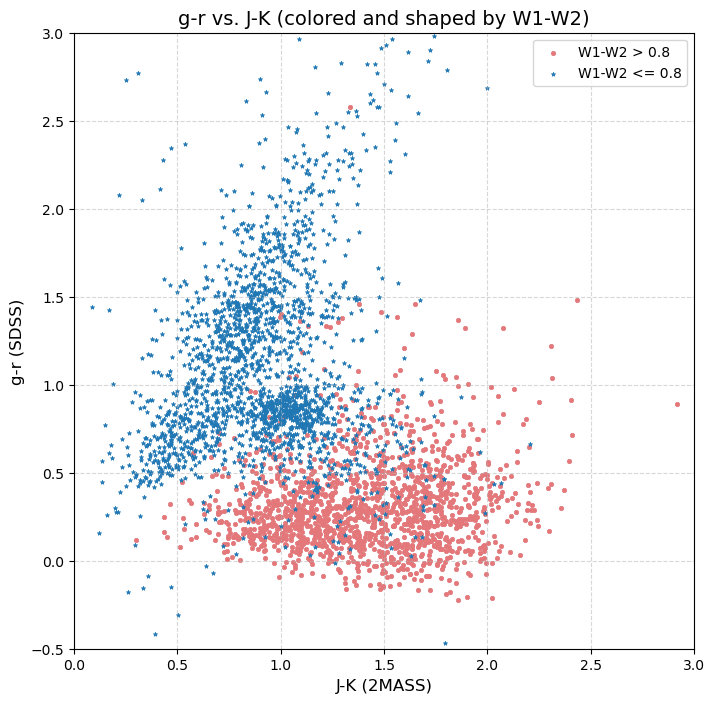}
\caption{ The $g - r$ versus\ $J - K$ color–color diagram of objects selected from
astrometric pre-selection.
Objects with $W1-W2 > 0.8$ (typically associated with quasars or AGNs) are shown as red circles, while those with $W1 - W2 \leq  0.8$ (likely stars and galaxies) are shown in blue.}
\label{fig:mid-ir}
\end{figure}

The distribution of the full zero-proper-motion sample in Fig.~\ref{fig:mid-ir} provides an empirical test of this interpretation. Sources with $W1-W2 \leq 0.8$ form a comparatively narrow, vertically extended sequence that broadly follows the predicted stellar locus. There is also a clear overdensity of points at the position expected for galaxies at around $J-K = 1.0$ and $g-r = 0.8$. In contrast, sources satisfying the commonly used mid-infrared AGN selection criterion, $W1-W2 > 0.8$, based on the criterion proposed by \citet{2012ApJ...753...30S}, are concentrated mainly in the lower (in $g-r$) and redder (in $J-K$) parts of the diagram, close to the regions occupied by the theoretical quasar tracks. The broad correspondence between the optical--near-infrared colour structure and the independent mid-infrared criterion therefore supports the main features predicted by the templates.

However, the separation is not complete. Sources on both sides of the $W1-W2=0.8$ threshold are present in the central and transitional parts of the diagram. The overlap is especially apparent in the regions where the stellar sequence approaches the blue and average quasar tracks, and where the red quasar tracks extend towards colours also occupied by stars, as shown by the theoretical colour tracks in Fig.~\ref{fig:colour}. Thus, neither the position in the $g-r$ versus $J-K$ diagram nor the $W1-W2$ criterion alone provides an unambiguous classification throughout the full colour space.

The comparison shows that some regions of the colour--colour diagram provide a more reliable separation between stars and quasars than others. The relatively blue-$J-K$ part of the stellar sequence remains dominated by sources with $W1-W2 \leq 0.8$, consistent with its interpretation as a comparatively clean stellar region. Similarly, the lower quasar locus contains a high concentration of sources with $W1-W2 > 0.8$ and can therefore be regarded as a relatively clean quasar region. Between these areas, the presence of both mid-infrared populations indicates that the transitional regions are more difficult to classify using photometric colours alone.

These results show that the theoretical tracks provide a useful first-order description of the source distribution, while the mid-infrared information offers an independent but imperfect empirical discriminator. Their combination identifies regions in which photometric classification is relatively reliable, as well as regions in which spectroscopic follow-up remains essential.

The spectroscopic classifications of the selected targets provide a more direct test of these regional interpretations and reveal a particularly notable difference between quasars found in the clean and transitional parts of the diagram.

\subsection{BAL quasars and peculiar quasars in the overlapping colour regions}

\begin{figure*}[htbp]
\centering
\includegraphics[width=0.48\textwidth]{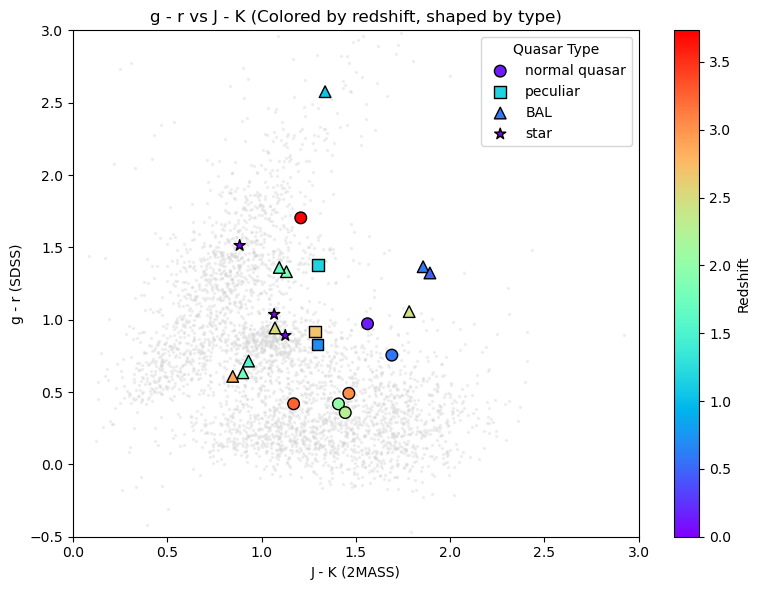}
\hfill
\includegraphics[width=0.48\textwidth]{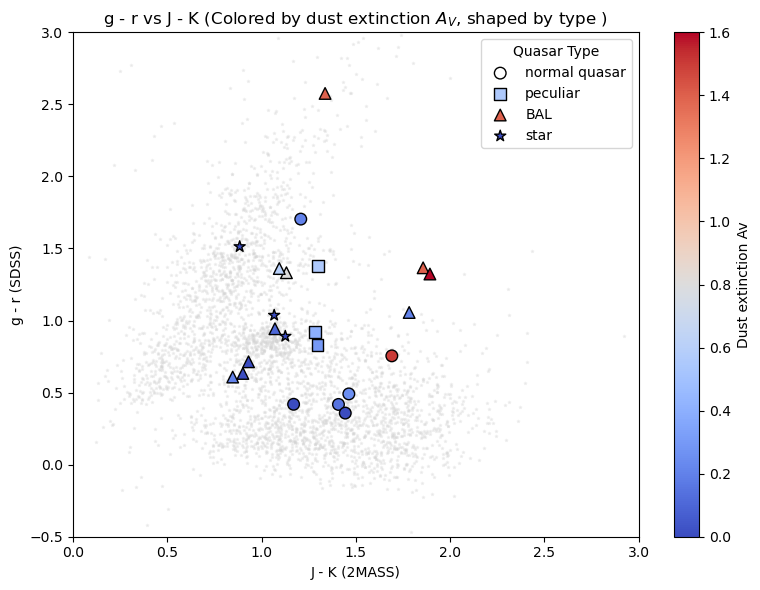}
\vspace{0.5em}

\includegraphics[width=0.48\textwidth]{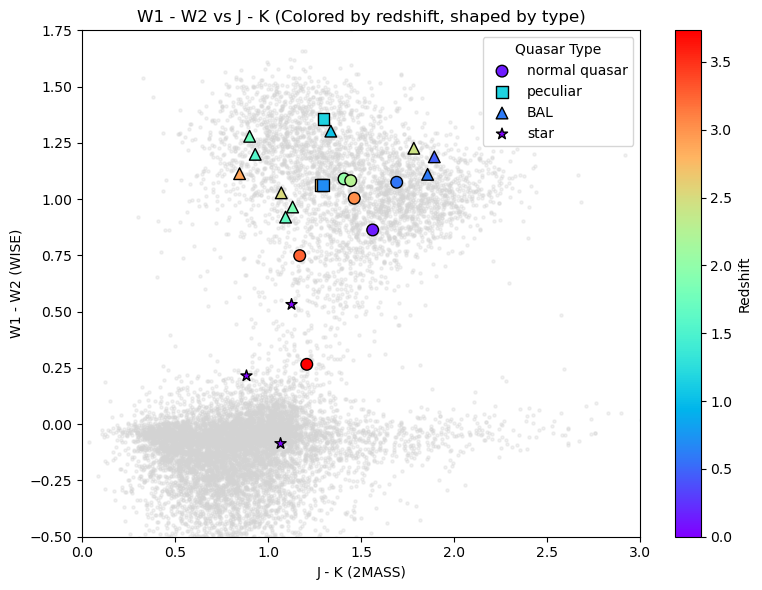}
\hfill
\includegraphics[width=0.48\textwidth]{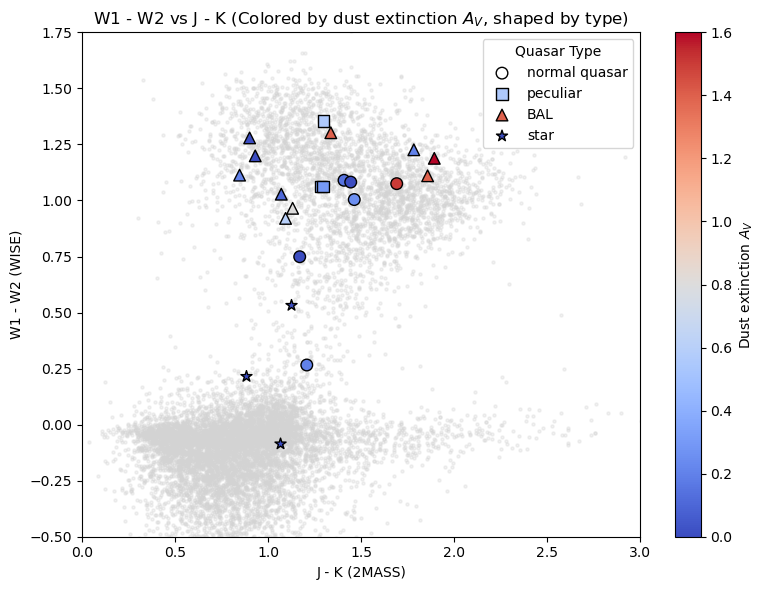}
\caption{Colour--colour diagrams showing the locations of the
spectroscopically observed targets. The top panels show $g-r$ versus $J-K$,
while the bottom panels show $W1-W2$ versus $J-K$. Marker shapes indicate
the spectroscopic classifications of the targets. In the left panels,
quasars are coloured according to their redshifts, while in the right
panels they are coloured according to their estimated dust extinctions,
$A_V$. Grey background points represent objects in the full
zero-proper-motion sample. Stars are shown using a fixed colour, since
redshift and quasar extinction estimates are not applicable to them.}
\label{fig:target_colours}
\end{figure*}

Figure~\ref{fig:target_colours} shows the positions of the observed targets
in the $g-r$ versus $J-K$ and $W1-W2$ versus $J-K$ diagrams, with marker shapes indicating their spectroscopic classifications. The quasars are
coloured by redshift in the left panels and by the estimated dust extinction, $A_V$, in the right panels. 

The spectroscopic classifications are broadly consistent with the expectations in the relatively well-separated regions of the colour--colour diagram. The single target observed in the star-safe region was confirmed to be a star, while all three targets located in the lower part of the traditional quasar-safe region in the $g-r$ versus $J-K$ diagram were confirmed as normal quasars. In contrast, the targets selected from the transitional regions show much more diverse classifications, including stars, normal quasars, BAL quasars, and peculiar quasars. Strikingly, only two of the 15 quasars confirmed in these more ambiguous regions are normal quasars. All BAL and peculiar quasars in our sample are located within these transitional regions.

No simple redshift sequence is apparent in either colour--colour plane. Quasars over a broad range of redshifts occupy overlapping regions, while
objects at similar redshifts can have substantially different colours. The displacement of BAL and peculiar quasars from the normal-quasar locus therefore cannot be explained by redshift alone. The panels coloured by
$A_V$ show that some of the reddest objects have relatively high estimated extinctions, suggesting that dust reddening contributes to their positions.

However, the relation between $A_V$ and broad-band colour is not one-to-one: quasars with comparable extinction values may occupy different
parts of the diagrams, and some objects with low fitted extinction also
show unusual colours. The explanation for this distribution is the complexity of the spectral features of the BAL and peculiar quasars. For BAL quasars, their broad line absorption features can suppress the flux of one or more photometric bands and therefore alter the observed broad band colours. This effect may shift some BAL quasars away from the relatively clean quasar region into the overlapping or transition parts of the colour--colour diagram. 

The peculiar quasars in our sample do not share a single spectral characteristic that can explain their locations in the colour--colour
diagrams. Instead, they exhibit a variety of departures from normal quasar spectra. These include a pronounced continuum break blueward of \ion{Mg}{2} in SDSS J214039+005705, unusually weak rest-frame ultraviolet emission lines in SDSS J225432+063826, and the absence of prominent \ion{Mg}{2} emission together with a continuum poorly reproduced by a reddened quasar template
in SDSS\,J235003+241943. Such variations in continuum shape and emission-line strength can affect the flux measured in individual photometric bands, producing unusual broad-band colours even when the fitted dust extinction
is modest. The locations of the peculiar quasars therefore likely reflect their heterogeneous spectral properties rather than a single effect such as redshift or dust reddening.

The distribution of the BAL quasars and the peculiar quasars in the photometrically ambiguous regions demonstrates the value of deliberately sampling outside the clean quasar locus. Selection methods optimized for high purity may avoid this regions because of their stronger stellar contamination, but may consequently miss quasars with complex spectral properties. The result of this work suggests that BAL quasars may occur at a relatively higher frequency than in the relatively clean quasar region. However, the present sample is small and was not selected to be statistically representative, so a larger sample is required to test this possibility.

\subsection{Unusual extinction properties in two low-redshift BAL quasars}

Among the BAL quasars in our sample, two objects, SDSS\,J182859+643725 and SDSS\,J231704+160807, exhibit spectral properties that are particularly unusual. In both cases, the observed flux at short wavelength falls systematically below the best fitting reddened quasar template. The SMC-like extinction curve cannot simultaneously reproduce the longer wavelength continuum and the suppression towards the blue.

\begin{figure}[htbp]
    \centering

    \includegraphics[width=\columnwidth]{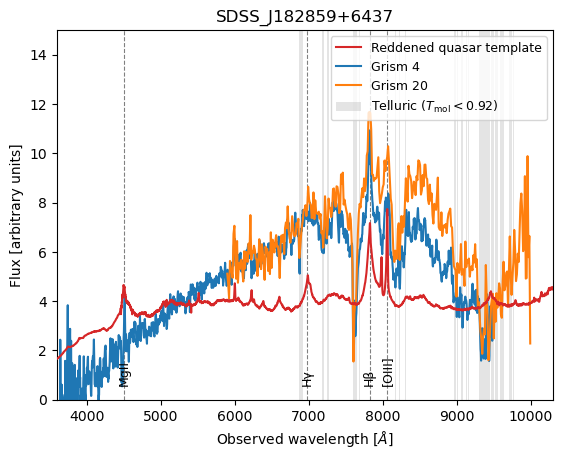}

    \vspace{0.5em}

    \includegraphics[width=\columnwidth]{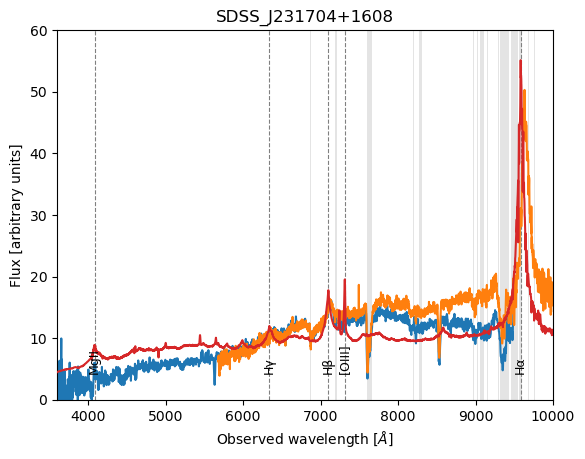}

    \caption{Observed spectra of SDSS J182859+643725 (top) and SDSS J231704+160807 (bottom), compared with their best-fitting reddened quasar templates. The blue and orange curves show the spectra obtained with ALFOSC Grism 4 and Grism 20, respectively, while the red curves represent the best-fitting reddened quasar templates scaled as in Fig.~\ref{fig:SDSS_J182859+643725} and Fig.~\ref{fig:SDSSJ231704+160807}, respectively. The vertical dashed lines indicate the expected observed wavelengths of the labeled spectral features. The grey shaded regions mark wavelength intervals affected by telluric absorption, where the atmospheric transmission is below 0.92. In both objects, the observed continua fall systematically below the templates at short wavelengths, whereas the most prominent absorption troughs occur mainly at longer observed wavelengths. In particular, the similar mismatch at the blue end of both spectra cannot readily be explained by the prominent absorption troughs alone and may indicate an extinction curve steeper than the SMC-like curve adopted in the fitting.}
    \label{fig:unusual_extinction}
\end{figure}

Part of this discrepancy may be related to the broad absorption features themselves. However, the strongest absorption features in these two spectra are mainly located at longer observed wavelengths, as shown in Fig.~\ref{fig:unusual_extinction}, whereas the shorter-wavelength continuum is systematically suppressed relative to the reddened template without comparably prominent absorption features. Their absorption features are located blueward of Balmer and Helium lines, which is unusual for BAL quasars \citep{2025ApJ...993..129L}.

One possible explanation is that the dust extinction towards these quasars differs from the SMC-like law adopted in the fitting. A steeper extinction curve would produce stronger attenuation at short wavelengths while leaving the longer-wavelength continuum comparatively less affected, and could therefore provide a better qualitative description of the observed spectral shape. A similarly steep extinction law has previously been suggested for reddened quasars for which an SMC-like curve could not reproduce both the optical spectra and the near-infrared photometry \citep{2013ApJS..204....6F}. Such a steep extinction curve may, in principle, be associated with a dust-grain population weighted towards smaller grains, which absorb and scatter short-wavelength radiation more efficiently.

\subsection{SPHEREx}

Finally we note that SPHEREx spectra are extremely useful for this kind of work as they provide extra features for 
redshift determination. In some case this was crucial for identifying the correct redshift (e.g., the case of SDSS\,J212319+254131).

\section{Conclusions and Outlook}
\label{conclusions}

We investigated 24 quasar candidates selected using Gaia astrometry and multi-wavelength photometry, with targets sampled from both relatively clean and transitional regions of the $g-r$ versus $J-K$ colour--colour diagram. Spectroscopic observations confirm 21 quasars and three stars. The quasars span $z=0.139$--$3.732$ and have estimated extinction values of $A_V=0$--$1.7$. Seven are classified as normal quasars (one of which is reminiscent of an LRD), 11 as BAL quasars, and three as peculiar quasars with weak emission lines.

The classifications in the relatively clean regions agree with the theoretical expectations: the single target selected from the stellar region was confirmed as a star, while all four targets selected from the safe quasar region were confirmed as normal quasars. In contrast, the transitional regions contain both stars and quasars, and all ten quasars confirmed there are BAL quasars. This suggests that BAL quasars may occur more frequently in photometrically ambiguous regions, although a larger and more systematic sample is required to test this possibility.

The addition of near-infrared spectra from SPHEREx was particularly useful in confirming redshifts when optical spectra were complex or difficult to interpret, including SDSS \, J182859+643725, SDSS \, J194143+610107, SDSS \, J212319+254131 and SDSS \, J214039+005705.

Two BAL quasars also show unusually suppressed blue continua that cannot be reproduced well by the adopted SMC-like reddened template. A steeper extinction curve may provide a possible explanation. 

Overall, our results show that combining astrometric selection, colour--colour sampling, and optical and near-infrared spectroscopy is effective for identifying quasars with unusual absorption and reddening properties.

Starting this year, the Purely Astrometric Quasar Survey \citep[PAQS,][]{2023Msngr.190...38K} 
will start building up a sample of of order 300.000 sources selected from Gaia astrometry down to a Gaia $G$ magnitude limit of 20.5 and followed up spectroscopically with the 4MOST instrument 
in Chile. This survey will sample the whole colour space (including the position of the stellar
track) and will hopefully give us a clean census of the quasar population down to the survey 
flux limit.

%As we have shown here a combination of astrometric pre-selection from Gaia matched with photometry from wide-area surveys makes it relatively easy to locate unusual and interesting quasars. Near-infrared observations from SPHEREx are also shown to be very useful for confirming redshifts in the cases where the optical spectra are complex and difficult to interprete (in particular cases like SDSS\,J182859+643725, SDSS\,J194143+610107, SDSS\,J212319+254131, and SDSS\,J214039+005705.

%%%%%%%%%%%%%%%%%%%%%%%%%%%%%%%%%%%%%%%%%%%%%%%%%%%%%%%%%%%%%%
\section*{Acknowledgements}
We thank Tom Reynolds for helpful advice about SPHEREx spectra and for helpful
discussions about the interpretation of several of the spectra. We thank Pascal Oesch for helpful
discussions concerning LRDs. 
This work made use of Astropy\footnote{\tiny{\url{https://www.astropy.org}}}: a community-developed core Python package and an ecosystem of tools and resources for astronomy \citep{2022ApJ...935..167A}.  

\paragraph*{Funding:}
The Cosmic Dawn Center (DAWN) is funded by the Danish National Research Foundation under grant DNRF140.  
JPUF is supported by the Independent Research Fund Denmark (DFF-4090-00079) and thanks the Carlsberg Foundation for support. JPUF acknowledges support by the European Research Council (ERC) under the European Union’s research and innovation program (ERC Grant HEAVYMETAL No.~101071865). 

\paragraph*{Data and materials availability:}
Based on observations made with the Nordic Optical Telescope, installed in the Spanish Observatorio del Roque de los Muchachos of the Instituto de Astrofísica de Canarias. The data presented here were obtained in part with ALFOSC, which is provided by the Instituto de Astrofisica de Andalucia (IAA) under a joint agreement with the University of Copenhagen and NOT. The NOT data were obtained under program 69-802.
This work has made use of data from the European Space Agency (ESA) mission Gaia, processed by the Gaia Data Processing and Analysis Consortium (DPAC). Funding for the DPAC has been provided by national institutions, in particular the institutions participating in the Gaia Multilateral Agreement.
This publication makes use of data products from the Spectro-Photometer for the History of the Universe, Epoch of Reionization and Ices Explorer (SPHEREx), which is a joint project of the Jet Propulsion Laboratory and the California Institute of Technology, and is funded by the National Aeronautics and Space Administration.

\bibliographystyle{aasjournal}
\bibliography{ref}

\end{document}

%% file: journals.tex
\def\apjl{ApJ}%
\def\apjs{ApJS}%
\def\ao{Appl.~Opt.}%
\def\aap{A\&A}%

%% file: notes_individual_objects.tex
The position of the objects in the colour-colour diagrams can be inspected in Fig.~\ref{fig:colour}

\subsubsection{SDSS\,J182859+643725 ($z = 0.609$)}

\begin{figure}[htbp]
\centering
\includegraphics[width=0.48\textwidth]{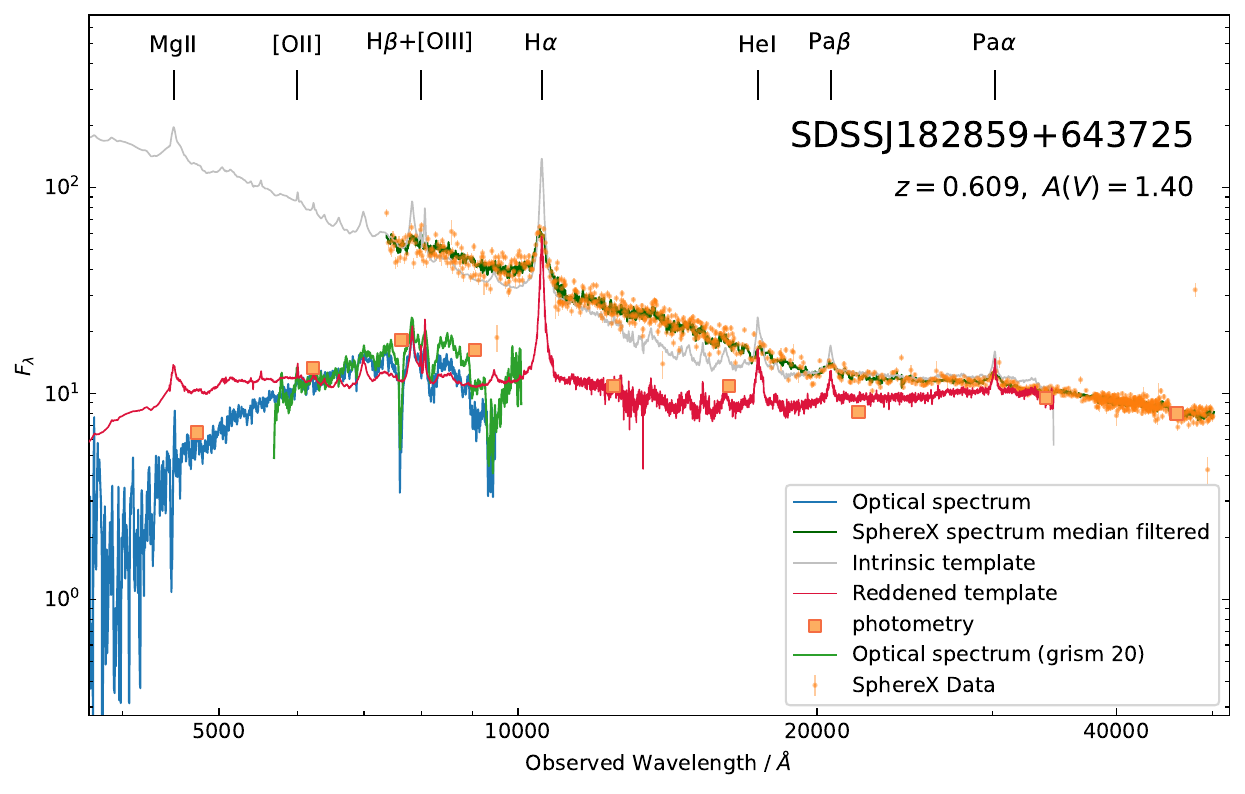}
\caption{The spectrum of SDSS\,J182859+643725. All figures of individual quasar spectra has
this layout: in blue (grism 4) and green (grism 20 when available) we show our optical spectra. The SPHEREx spectra are shown with orange (individual points) and green (median filtered). The photometry from SDSS, 2MASS, and WISE is shown with orange squares. The quasar template is shown in grey (unreddened) and red (reddened with an SMC extinction curve). The redshift and the assumed $A_\mathrm{V}$ is indicated on the figure. We classify this source as a peculiar BAL quasar. The SPHEREx spectrum is blended 
with light from a brighter source.}
\label{fig:SDSS_J182859+643725}
\end{figure}

This object has colours $g-r = 1.37$, $J-K=1.86$, and $W1-W2 = 1.11$. It was selected
due to its red optical and near-infrared colours.
For this quasar we identify a redshift of $z = 0.609$, determined based on a
prominent broad emission feature near 7800 {\AA}, which is interpreted as H$\beta$ (Fig.~\ref{fig:SDSS_J182859+643725}). 
This redshift is consistent with the SPHEREx spectrum, which shows very strong Hydrogen 
emission lines consistent with this redshift (H$\alpha$, P$\alpha$, P$\beta$). The SPHEREx spectrum has the wrong
continuum shape compared to the photometry, which is due to contamination of emission from another
brighter source within the SPHEREx PSF.
A dust extinction value of A$_\mathrm{V}$ = 1.4 is estimated, as it yields a good match between
the photometric data and the quasar template in the near-infrared region. This
dust extinction value shows that the quasar spectrum is heavily reddened by dust.

There is a strong absorption trough on the red side of the H$\beta$ emission, above what can 
be accounted for by telluric absorption at these observed wavelengths, which we have not 
been able to identify. One possibility is \ion{He}{1}. There is also evidence for blueshifted
H$\beta$ absorption, but this is blended with the telluric A band. There is also 
blueshifted \ion{Mg}{2} absorption.

%The
%spectrum shows mixed absorption troughs in 6300 -- 9000 {\AA}, with multiple
%sharp features attributed to low-ionization metal lines, including \ION{Fe}{I}, \ION{Ca}{II}, 
%and \ion{Na}{i},
%and rarer neutral species such as \ion{Mn}{i}, \ion{Ti}{i}, and \ion{Cr}{i}. 

At shorter wavelengths, the
observed spectrum falls well below the reddened template. This discrepancy
may indicate that the actual dust extinction curve is steeper in the UV than the
adopted SMC-like law, which is discussed in \citet{2013ApJS..204....6F}.
Due to the peculiar nature of this spectrum, a second observation was carried
out using grism 20, which extends the wavelength coverage toward longer
wavelengths. The result from this second observation is consistent with the
grism 4 spectrum in the overlapping region, confirming the reliability of the
earlier data. In the newly covered longer-wavelength range, abroad absorption
feature is observed just before H$\alpha$, which supports the classification of this
object as a peculiar BAL quasar. We subsequently realized that this quasar is also observed
as part of BOSS, but as far as we can see the redshift has not been established prior to 
our study. The source is detected by LOFAR with a flux of 8.20$\pm$0.08 mJy.

\subsubsection{SDSS\,J194143+610107 ($z = 1.042$)}
\begin{figure}[htbp]
\centering
\includegraphics[width=0.48\textwidth]{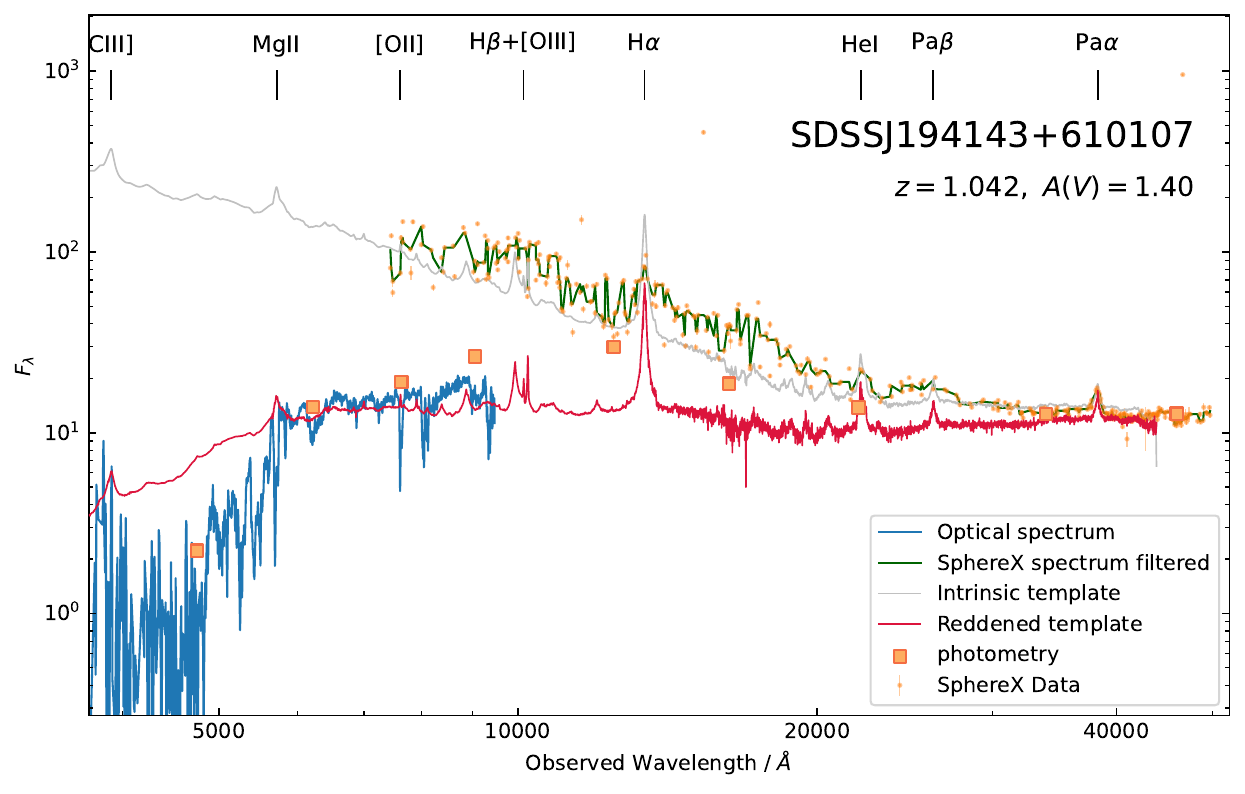}
\caption{The spectrum of SDSS\,J194143+610107. We classify this source as a FeLoBAL quasar. 
The SPHEREx spectrum is blended with light from a brighter source.}
\label{fig:SDSS_J194143+610107}
\end{figure}

This object has colours $g-r = 2.58$, $J-K=1.34$, and $W1-W2 = 1.30$. It was selected
for follow-up spectroscopy due to its very red optical colour (the reddest in the sample).
This quasar spectrum (Fig.~\ref{fig:SDSS_J194143+610107}) exhibits a heavily suppressed continuum, with no prominent
$z = 1.042$ based on the identification of several strong absorption lines,
including \ion{Fe}{2}$\lambda$3301,$\lambda$3363 and CaII($\lambda$3934,$\lambda$3969). At this redshift
other absorption features align well with some low-ionization transitions such
as \ion{Ti}{2} and \ion{Mg}{1}. The quasar shows broad and deep absorption troughs across
the 6000 to 8500 {\AA} range, and many absorption features are blended, forming
a complex profile. The dust extinction is estimated to be A$_\mathrm{V}$ = 1.4, which gives
a good match between the template and the photometric data in the NIR. In the 
SPHEREx spectrum there are emission lines consistent with Pa$\alpha$ and H$\alpha$ at the proposed redshift.
The SPHEREx spectrum is contaminated by another bright source at the edge of the SPHEREx PSF,
which explains why the slope does not match the photometry.
The combination of multiple low-ionization metal lines (e.g., \ion{Fe}{2}, \ion{Mg}{1}, \ion{Ti}{2}),
suppressed UV emission lines, strong continuum depression, and wide absorption
features lead us to classify this object as a FeLoBAL quasar. We subsequently realised 
that this object has also been observed as part of BOSS, but as far as we can see the redshift 
has not been established prior to 
our study. The source is detected by LOFAR with a flux of 
2.23$\pm$0.12 mJy.

\subsubsection{SDSS\,J203008+15241616 ($z = 1.741$)}
\begin{figure}[htbp]
\centering
\includegraphics[width=0.48\textwidth]{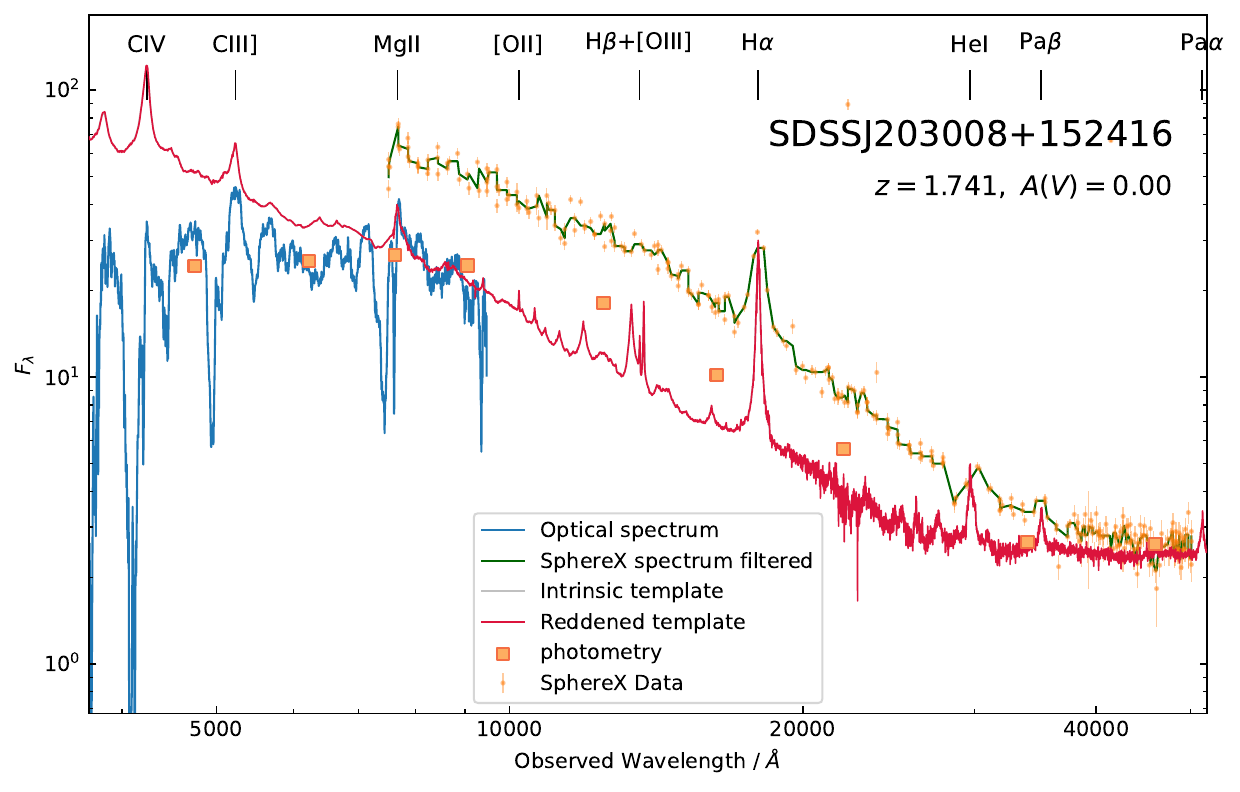}
\caption{The spectrum of SDSS\,J203008+152416. We classify this source as a BAL quasar.}
\label{fig:SDSS_J203008+152416}
\end{figure}

This object has colours $g-r = 0.631$, $J-K=0.90$, and $W1-W2 = 1.28$. It was selected
due to be redder than typical quasars.
This quasar exhibits broad absorption features blueward of the \ion{C}{4} and \ion{C}{3}]
emission lines, accompanied by significant suppression of the continuum in
the rest-frame UV region (Fig.~\ref{fig:SDSS_J203008+152416}). The redshift is determined based on a prominent
emission feature near 5230 {\AA}, identified as \ion{C}{3}] $\lambda$1909. This is consistent 
with the presence of H$\alpha$ emission in the expected wavelength range in the SPHEREx
spectrum. The SPHEREx spectrum is again contaminated by light from another source. There is no
evidence of significant dust reddening (A$_\mathrm{V}$ = 0).
The spectrum reveals prominent BAL characteristics. Therefore we label this
object as a BAL quasar.

\subsubsection{SDSS\,J204757+002431 ($z = 1.850$)}
This object has colours $g-r = 1.33$, $J-K=1.13$, and $W1-W2 = 0.97$. It was selected
due to its red optical colour.
The redshift of this quasar is determined primarily from a set of strong absorption features, as its emission lines are significantly suppressed. In particular,
\ion{Fe}{2} $\lambda$2374, \ion{Fe}{2} $\lambda$2382, \ion{Mn}{2} $\lambda$2576, and \ion{Mn}{2} $\lambda$2606 appear prominently
in the observed spectrum. Consistent with $z = 1.850$, we also find that emission
features such as \ion{C}{4}, \ion{C}{3}] and \ion{Mg}{2} align well with the template. The 
SPHEREx spectrum shows the presence of an H$\alpha$ line consistent with this redshift. 
The spectrum exhibits several defining characteristics of a FeLoBAL quasar, including broad and blended absorption troughs, extensive low-ionization metal
absorption (\ion{Fe}{2}, \ion{Mn}{2}, and \ion{Mg}{2}) and continuum suppression over a wide
wavelength range. The spectral shape is similar to the known FeLoBAL SDSS
J090152.05+624342.6 \citep{2017FrASS...4...40W}. The position is covered by the FIRST survey,
but no signal is detected from the source. We label this object as a FeLoBAL quasar.

\begin{figure}[htbp]
\centering
\includegraphics[width=0.48\textwidth]{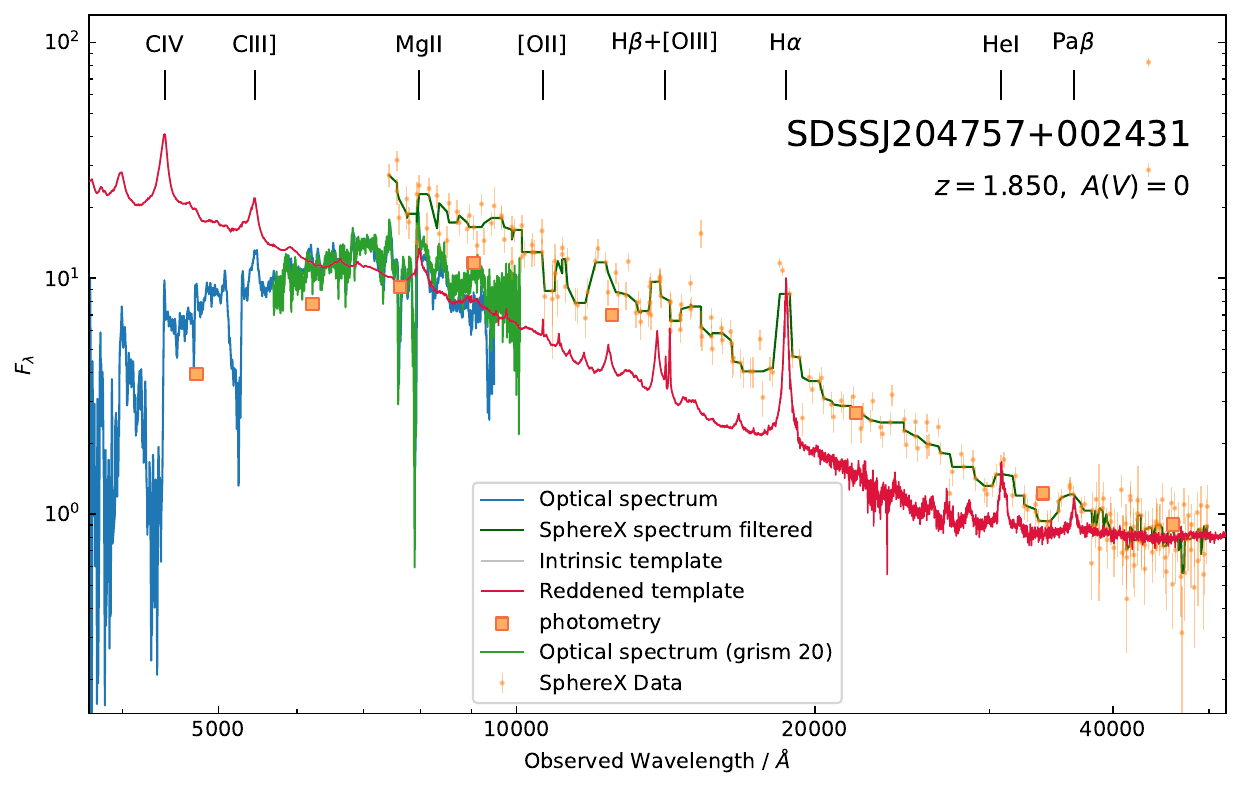}
\caption{The spectrum of SDSS\,J204757+002031. We classify this source as a FeLoBAL quasar.}
\label{fig:SDSS_J204757+002031}
\end{figure}

\subsubsection{SDSS\,J210215+230730 ($z=0.139$)}
\begin{figure}[htbp]
\centering
\includegraphics[width=0.48\textwidth]{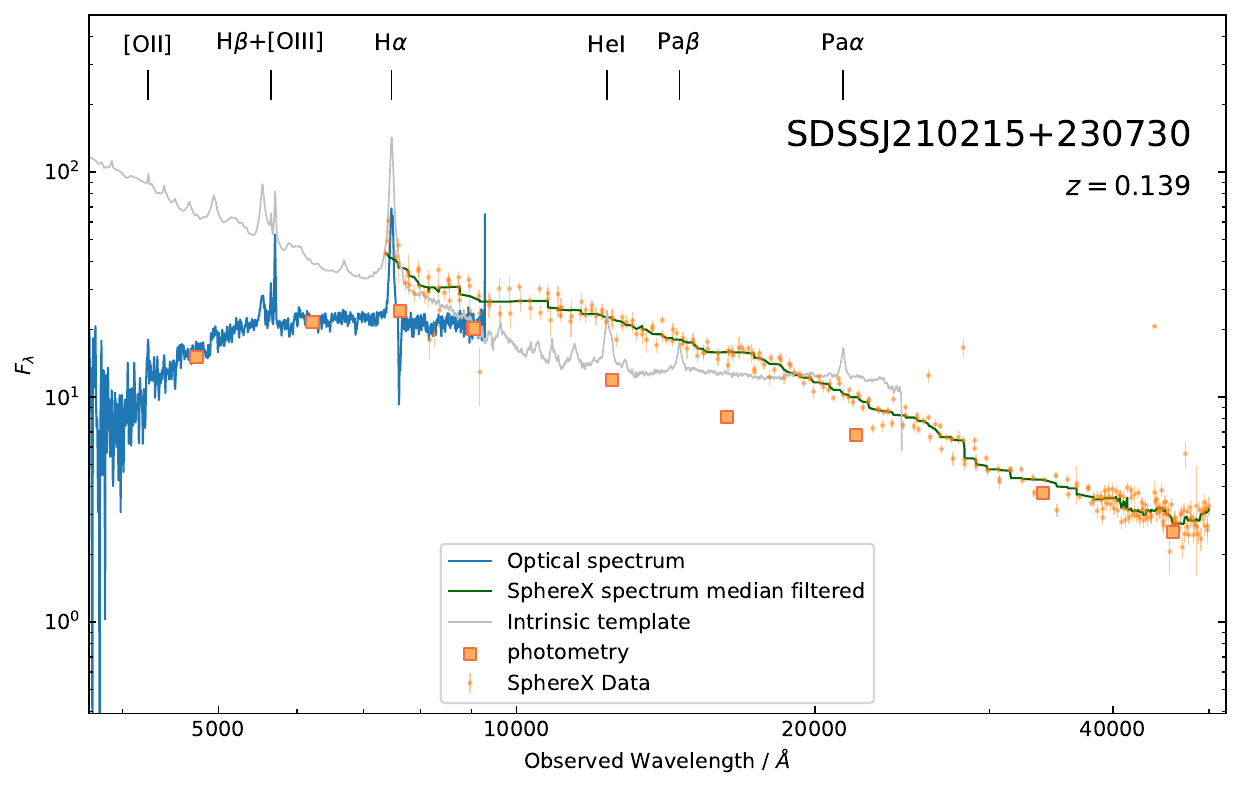}
\caption{The spectrum of SDSS\,J210215+230730. We classify this source as a normal quasar. It is somewhat reminiscent of an LRD. This is the lowest redshift source in the sample. We do not overplot a reddened quasar template as it matches very poorly with any
extinction. 
}
\label{fig:SDSS_J210215+230730}
\end{figure}

This object has colours
$g-r = 0.97$, 
$J-K = 1.56$, and
$W1-W2 = 0.86$. 
It was selected for spectroscopic follow-up due to its red optical and near-infrared
colours. We determine a low redshift of 0.139 based on broad H$\alpha$ and narrow
[\ion{O}{2}] and [\ion{O}{3}] emission (see Fig.~\ref{fig:SDSS_J210215+230730}). 
H$\beta$ and \ion{Mg}{2} is also detected. The continuum is very red. In Fig.~\ref{fig:SDSS_J210215+230730} we have not
plotted the quasar template as it is not possible to match the spectrum with a quasar 
template reddened by SMC-like extinction. 
In SPHEREx the shape of the spectrum matches the photometry fairly well, but the 
photometry falls below the SPHEREx measurements, probably due to variability. The SPHEREx measurements 
are not contaminated by light from other sources. In SDSS the object is extended
and is classified as a galaxy, but the $W1-W2$ colour is inconsistent with a galaxy (or 
stellar) spectrum. This is an object with contributions to the 
spectrum from both a central obscured AGN and from the underlying host galaxy, somewhat reminiscent of Little Red Dots at higher redshifts \citep[LRDs,][]{2024ApJ...963..129M, 2026ApJ..1003..234R}.
The shape of the continuum is similar, but the H$\alpha$ line does not have 
the exponential profile seen in LRD spectra. 
We classify it as a normal quasar/AGN due to its broad H$\alpha$ line.  

\subsubsection{SDSS\,J210558+222457 (star)}
This object has colours $g-r = 1.04$, $J-K=1.07$, and $W1-W2 = -0.09$. It was selected
due to its red optical and near-infrared colours and low $W1-W2$ colour suggesting it could 
be at high redshift.
The spectrum of this object shows a distinctly stellar shape, lacking any broad
emission lines characteristic of quasars. The spectrum is well matched by stars of 
spectral type K.

\begin{figure}[htbp]
\centering
\includegraphics[width=0.48\textwidth]{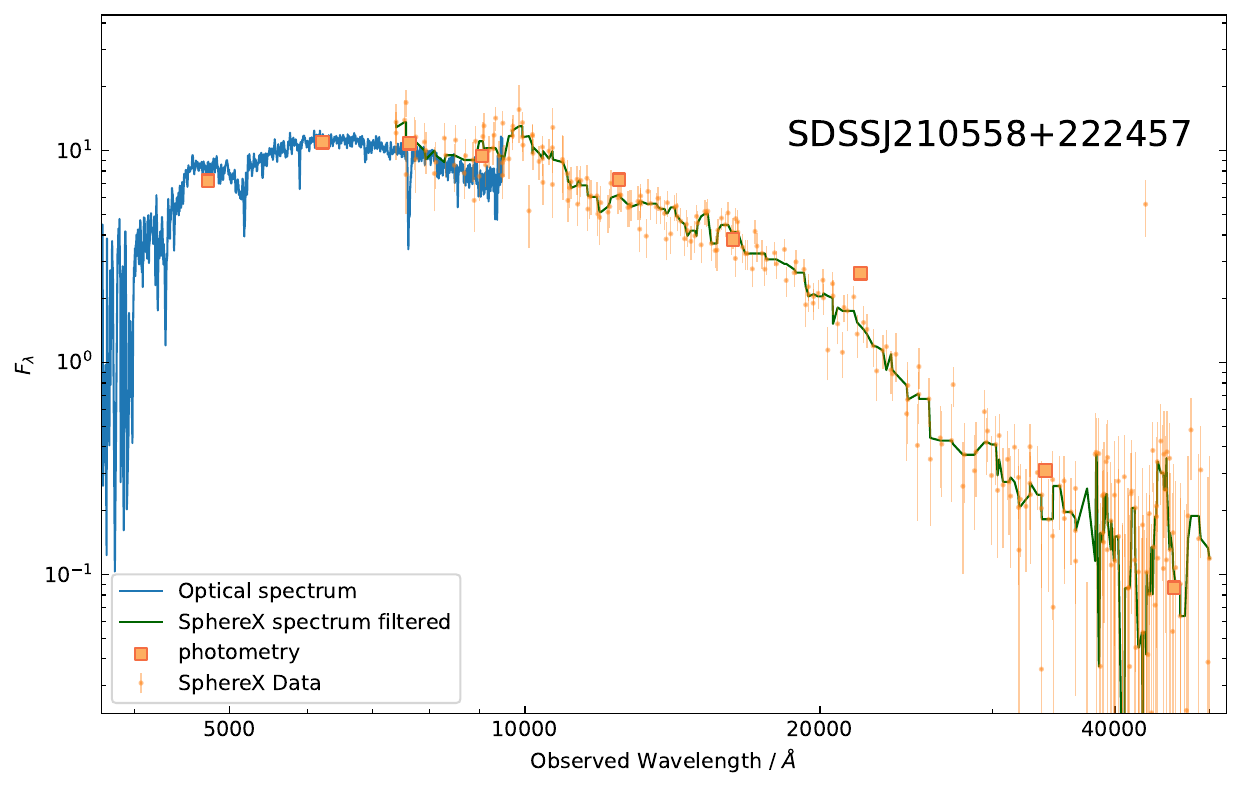}
\caption{The spectrum of SDSS\,J210558+222457. We classify this source as a K-type star.}
\label{fig:SDSSJ210558+222457}
\end{figure}

\subsubsection{SDSS\,J212319+254131 ($z = 2.520$)}
\begin{figure}[htbp]
\centering
\includegraphics[width=0.48\textwidth]{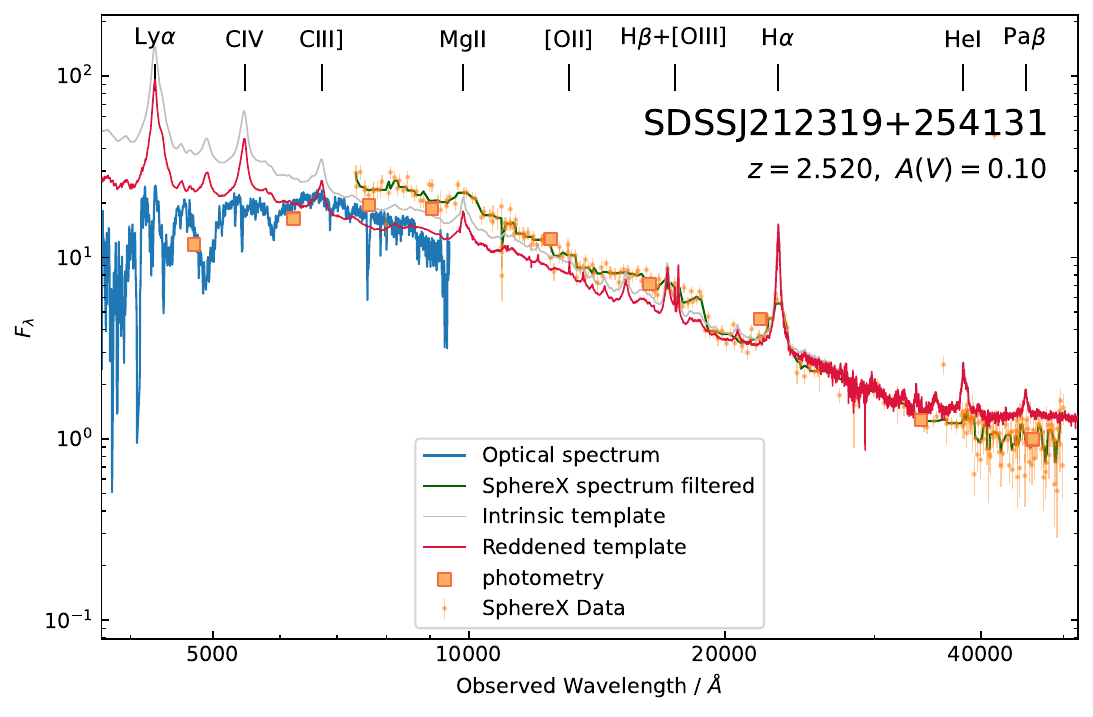}
\caption{The spectrum of SDSS\,J212319+254131. We classify this source as a BAL quasar.}
\label{fig:SDSSJ212319+254131}
\end{figure}

This object has colours $g-r = 0.94$, $J-K=1.07$ and $W1-W2 = 1.03$. It was selected
due to its red optical colour.
Similar to SDSSJ\,194143+610107 and SDSSJ\,204757+002431, this quasar exhibits
no prominent emission features in the optical (see Fig.~\ref{fig:SDSSJ212319+254131}), making redshift determination in the optical relying on
absorption lines. At $z = 2.36$, a series of \ion{Fe}{2} absorption features (e.g., $\lambda$2344,
$\lambda$2374, $\lambda$2382) and low-ionization lines such as \ion{Mg}{2} $\lambda$2796 $\lambda$2803.
There is a also a broad BAL-line in the blue end of the spectrum consistent with \ion{C}{4}
at $z=2.36$.
In the SPHEREx spectrum, however, the observed spectrum shows a prominent H$\alpha$ emission feature, 
corresponding to a redshift of $z = 2.52$ (Fig.~\ref{fig:SDSSJ212319+254131}). Taken together,
this shows that this quasar is a high-velocity detached BAL quasar similar to the quasars studied by 
\citet{2020A&A...634A.111F} and \citet{2026arXiv260701330M}. It is however, unusual to see 
detached low-ionization lines.

\subsubsection{SDSS\,J213510+272852 ($z = 2.430$)}
\begin{figure}[htbp]
\centering
\includegraphics[width=0.48\textwidth]{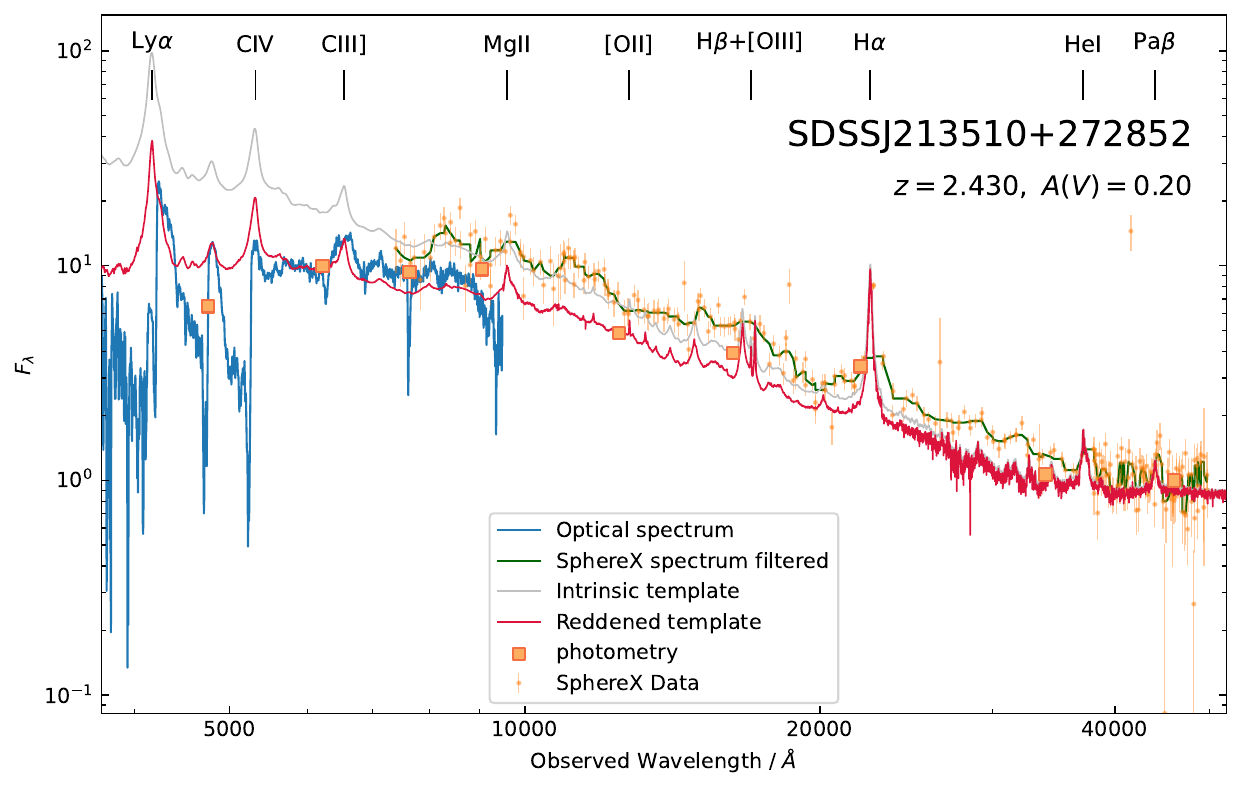}
\caption{The spectrum of SDSS\,J213510+272852. We classify this source as a BAL quasar.}
\label{fig:SDSSJ213510+272852}
\end{figure}

This object has colours $g-r = 1.06$, $J-K=1.78$, and $W1-W2 = 1.23$. It was selected
for follow-up spectroscopy due to its red optical colour.
The redshift of this quasar is determined based on the observed \ion{C}{3}] emission
feature, with supporting emission from Ly$\alpha$ and \ion{C}{4} clearly visible in the
spectrum (see Fig.~\ref{fig:SDSSJ213510+272852}). There are clear BAL features bluewards
of the \ion{C}{4}, \ion{Si}{4} and Ly$\alpha$ emission lines. In addition to the high-ionization features, 
absorption from the low-ionization \ion{Mg}{2} doublet is present, allowing a further
classification as a low-ionization BAL (LoBAL) quasar. A reddening value of
A$_\mathrm{V}$ = 0.2 provides an excellent match between the template and photometry.
The source is detected by LOFAR with a flux of 6.27$\pm$0.41 mJy.
We label this quasar as a LoBAL quasar.

\subsubsection{SDSS\,J214039+005705 ($z = 1.156$)}
\begin{figure}[htbp]
\centering
\includegraphics[width=0.48\textwidth]{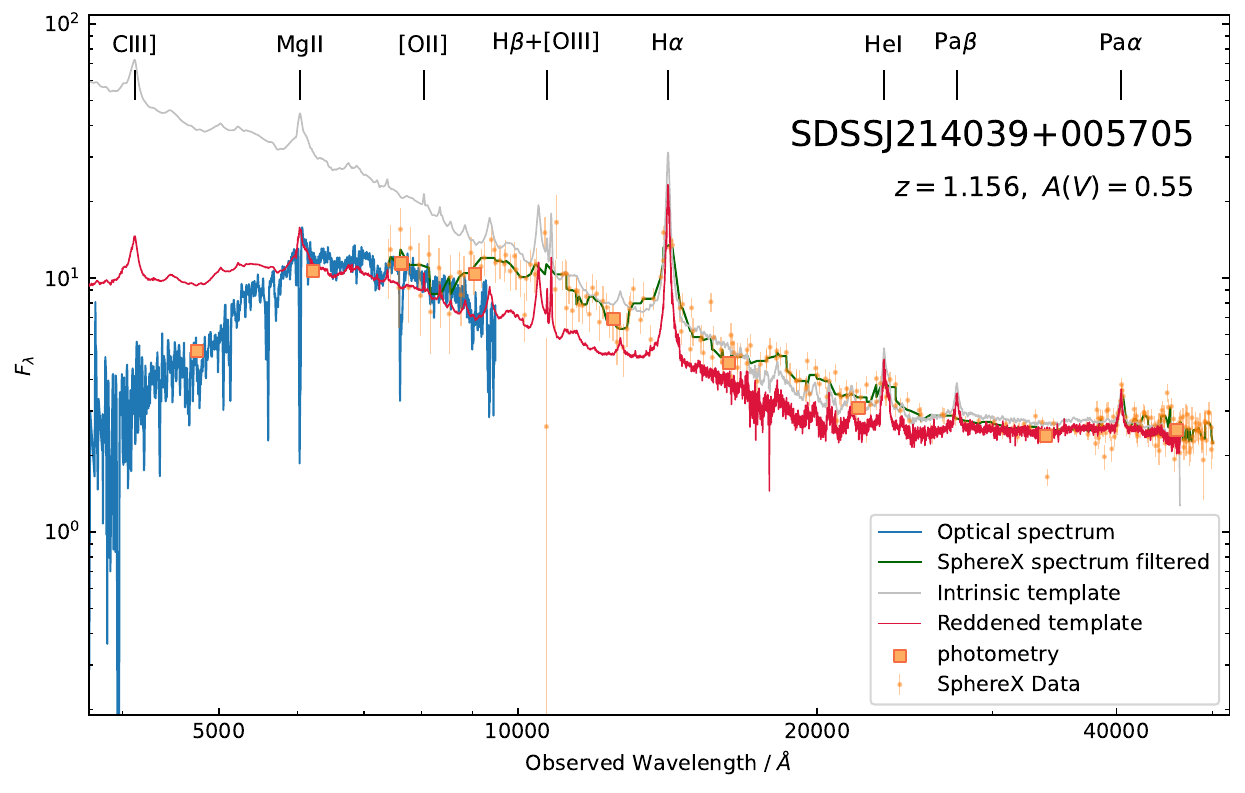}
\caption{The spectrum of SDSS\,J214039+005705. We classify this source as a peculiar quasar.}
\label{fig:SDSS214039+005705}
\end{figure}

This object has colours $g-r = 1.38$, $J-K=1.30$, and $W1-W2 = 1.36$. It was selected
for follow-up spectroscopy due to its red optical colour.
The redshift of this quasar is determined to be $z = 1.1555$, based on the alignment 
of several prominent absorption features from Fe, Mn, and Mg (see Fig.~\ref{fig:SDSS214039+005705}). 
No strong emission lines are
clearly identifiable in the optical. The continuum breaks blueward of Magnesium. 
In the SPHEREx spectrum there are emission lines from H$\alpha$, \ion{He}{1}, and Pa$\alpha$
consistent with $z=1.1555$.
Over most of the spectra range a normal quasar template reddened by SMC-like extinction
with A$_\mathrm{V}=0.55$ gives a decent match, but below about 3000 {\AA} the continuum
seems to break. The position is covered by the FIRST survey,
but no signal is detected from the source. We classify this source as a peculiar quasar.

\subsubsection{SDSS\,J215731+183120 ($z = 2.020$)}
This object has colours $g-r = 0.42$, $J-K=1.41$, and $W1-W2 = 1.09$. It was selected
to have colours typical of a normal $z<2.8$ quasar.
The redshift of this quasar is estimated to be $z = 2.020$, primarily based on the
alignment of prominent broad emission features in the optical including \ion{C}{3}], \ion{C}{4}, and Ly$\alpha$.
The SPHEREx spectrum shows emission from H$\alpha$, H$\beta$, and Pa$\beta$. 
The extinction value A$_\mathrm{V}$ = 0.15 provides a good match between the observed
spectrum and the reddened template, indicating that the quasar is mildly
reddened. The spectrum shows no clear signs of broad absorption troughs
or strong low-ionization absorption lines. 
The source is detected by LOFAR with a flux of 2.57$\pm$0.10 mJy.
Therefore, we classify SDSS\,J2157+183120 as a normal quasar.

\begin{figure}[htbp]
\centering
\includegraphics[width=0.48\textwidth]{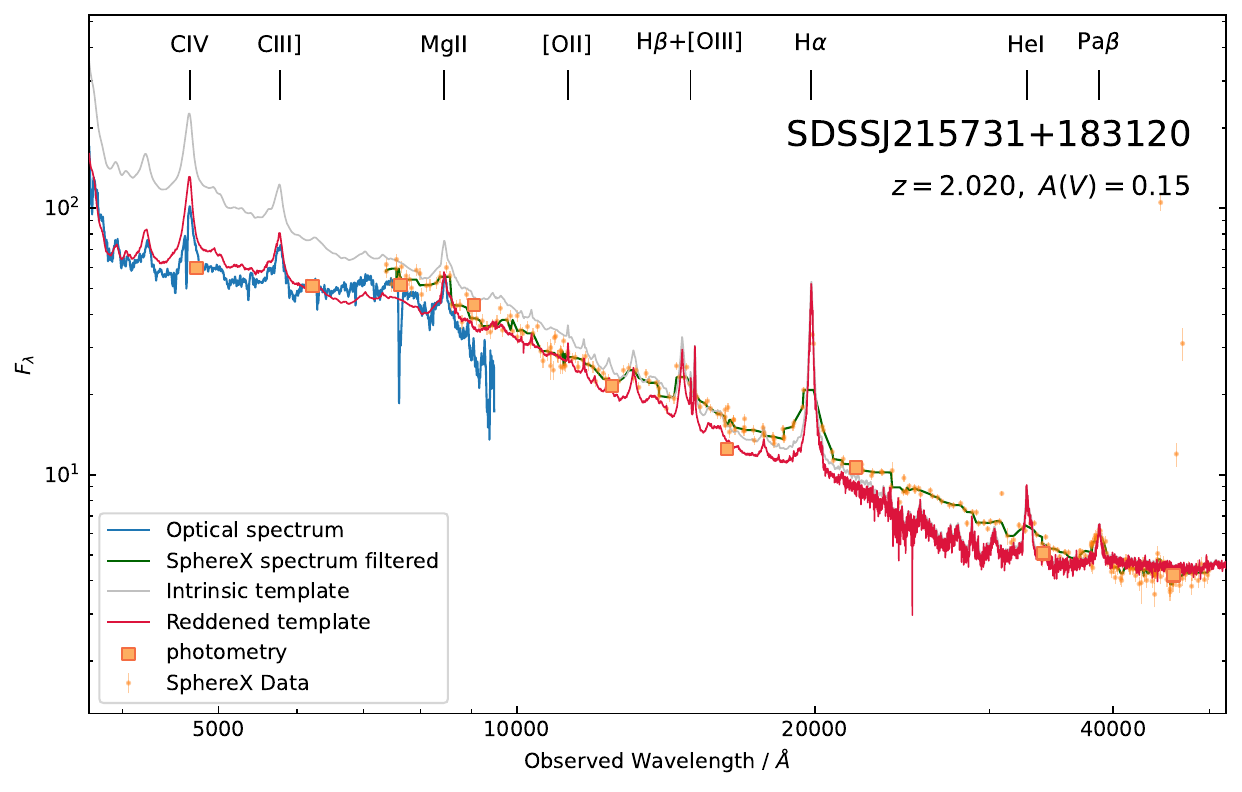}
\caption{The spectrum of SDSS\,J215731+183120. We classify this source as a normal quasar.}
\label{fig:SDSS215731+183120}
\end{figure}

\subsubsection{SDSS\,J215856+285434 ($z=1.553$)}
\begin{figure}[htbp]
\centering
\includegraphics[width=0.48\textwidth]{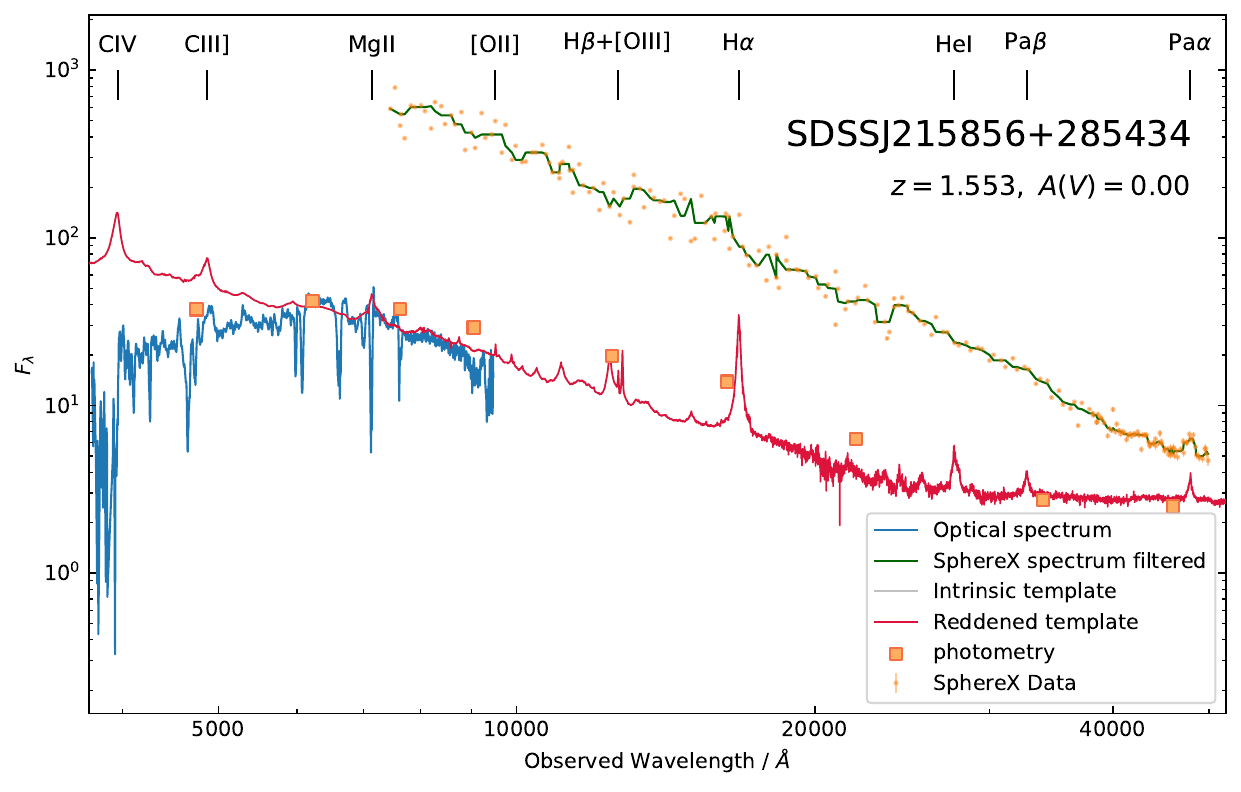}
\caption{The spectrum of SDSS\,J215856+285434. We classify this source as a BAL quasar. The SPHEREx spectrum is blended 
with light from a brighter source.}
\label{fig:SDSS215856+285434}
\end{figure}

This object has colours $g-r = 0.71$, $J-K=0.93$, and $W1-W2 = 1.20$. It was selected
for follow-up spectroscopy due to its mildly red optical colour compared to normal quasars.
The redshift of this quasar is estimated to be $z=1.553$ based on the identification of broad absorption features, 
including \ion{Fe}{2} $\lambda$2344, $\lambda$2374, $\lambda$2382 and
\ion{Mg}{2}$\lambda$2586, $\lambda$2600, $\lambda$2802 (see Fig.~\ref{fig:SDSS215856+285434}). 
The presence of strong, wide, and overlapping
low-ionization absorption features—especially those from \ion{Fe}{2} supports its classification as a low-ionization BAL quasar. 
Once the redshift is determined, we also
find that the emission lines of \ion{C}{4}, \ion{C}{3}], and \ion{Mg}{2} align well with the template, 
although all are significantly suppressed. The SPHEREx spectrum is contaminated by light from a much brighter source. 
Despite of this, Pa$\alpha$ emission is detected from the quasar at the expected wavelength. 

\subsubsection{SDSS\,J221036+282050 ($z=3.732$)}

\begin{figure}[htbp]
\centering
\includegraphics[width=0.48\textwidth]{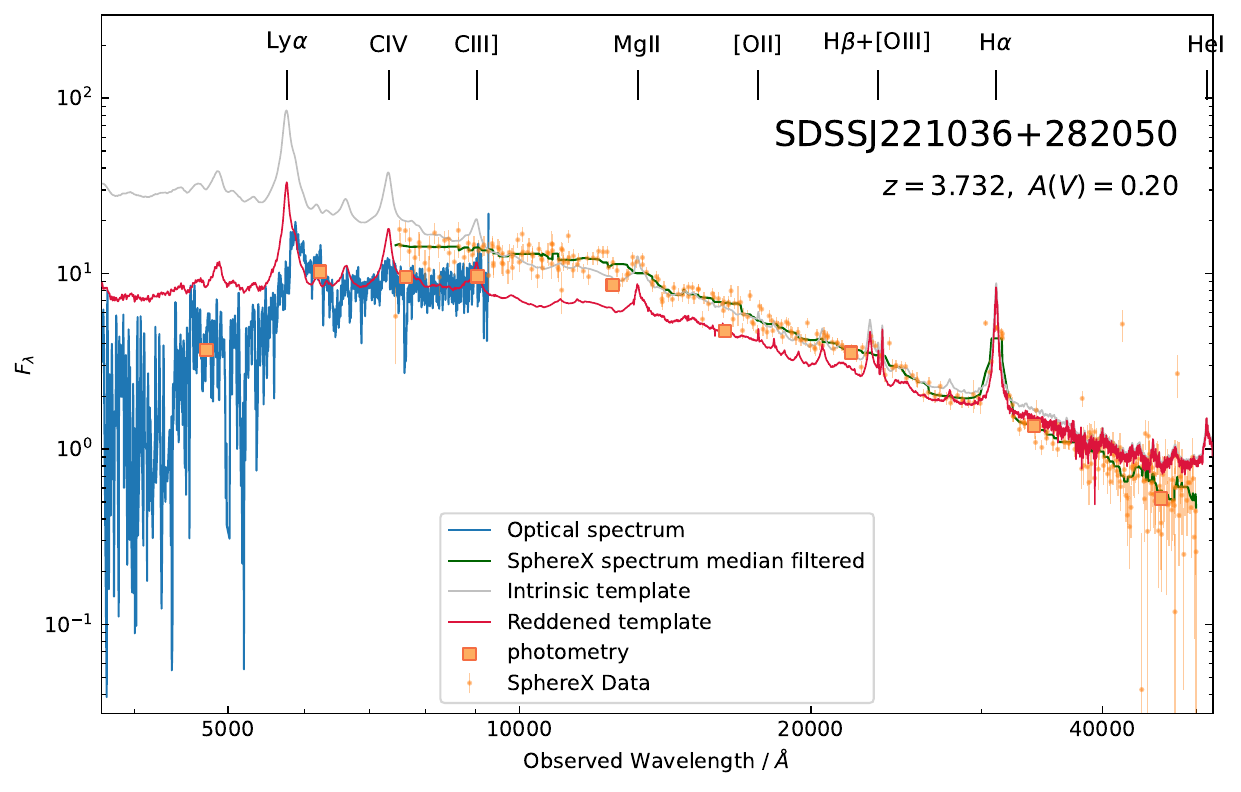}
\caption{The spectrum of SDSS\,J221036+282050. We classify this source as a normal quasar. This is the highest redshift source in the sample.}
\label{fig:SDSS221036+282050}
\end{figure}

This object has colours 
$g-r = 1.70$, 
$J-K = 1.21$, and 
$W1-W2 = 0.27$.
It was selected for follow-up spectroscopy due to its red optical colour and 
the low $W1-W2$ colour, which could suggest a redshift larger than 2.8. 
We determine a redshift of $z=3.732$ based on weak Ly$\alpha$ in the optical
and H$\alpha$ in the SPHEREx spectrum (see Fig.~\ref{fig:SDSS221036+282050}). 
The spectrum is well matched with a composite quasar spectrum 
mildly reddened ($A_\mathrm{V} = 0.2$ mag). We classify this as a normal quasar.

\subsubsection{SDSS\,J222839+172759 ($z=2.932$)}

This object has colours
$g-r = 0.39$,
$J-K = 1.37$, and
$W1-W2 = 0.78$.
It was selected for follow-up spectroscopy due to its blue optical colour 
and red near-infrared colours. 
We determine a redshift of $z=2.932$ based on Ly$\alpha$ and \ion{C}{3}]
in the optical and H$\alpha$ in the SPHEREx spectrum. The spectrum is mildly
reddened. There is a broad and detected BAL-feature from \ion{C}{4}.
We categorize the source as a BAL quasar. 

\begin{figure}[htbp]
\centering
\includegraphics[width=0.48\textwidth]{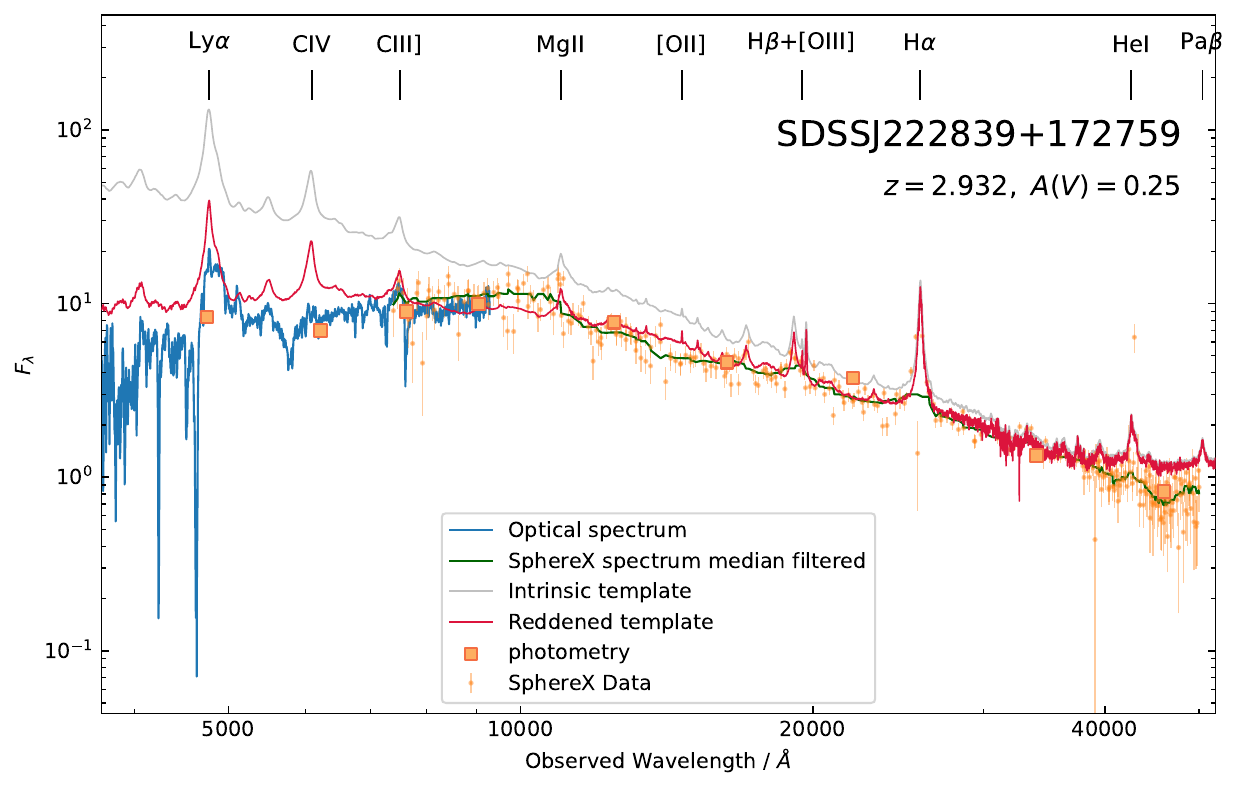}
\caption{The spectrum of SDSS\,J222839+172759. We classify this source as a BAL quasar.}
\label{fig:SDSS222839+172759}
\end{figure}

\subsubsection{SDSS\,J222934+101907 (star)}
\begin{figure}[htbp]
\centering
\includegraphics[width=0.48\textwidth]{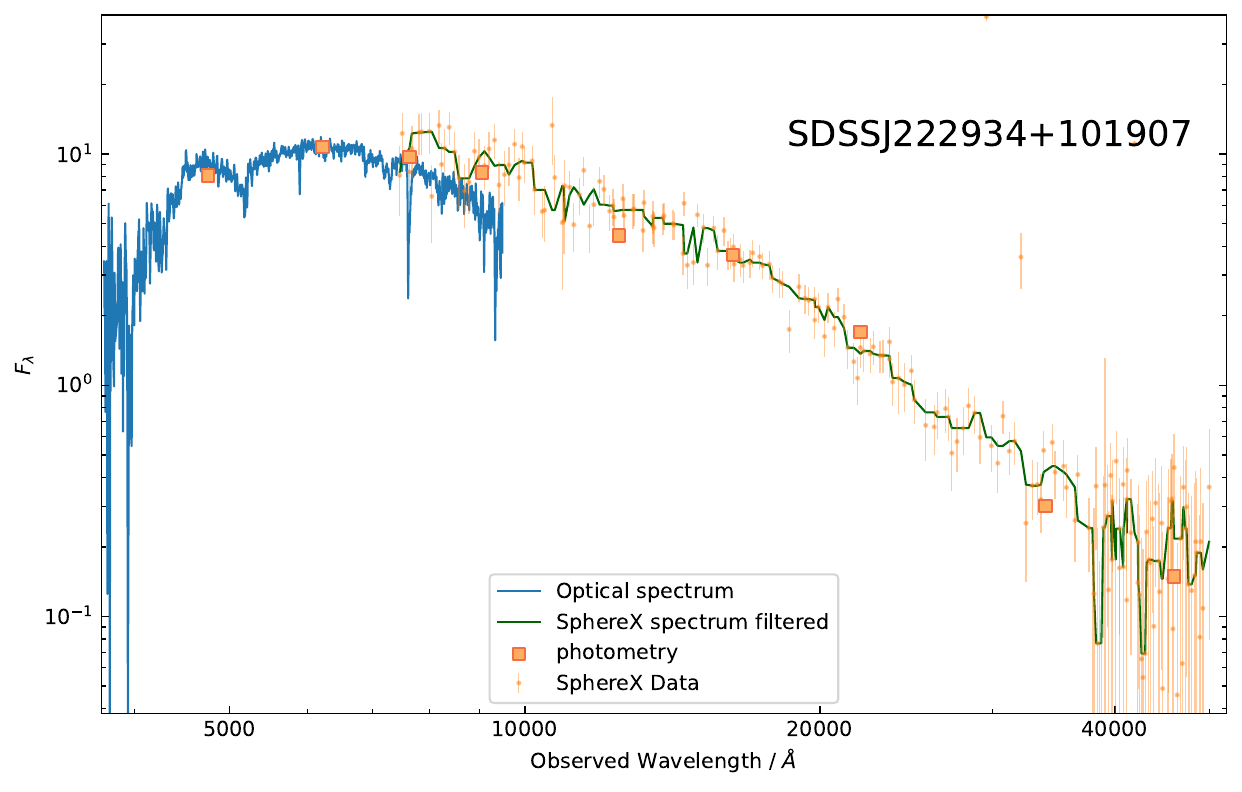}
\caption{The spectrum of SDSS\,J222934+101907. We classify this source as K-type star.}
\label{fig:SDSS222934+101907}
\end{figure}

This object has colours $g-r = 0.89$, $J-K=1.13$, and $W1-W2 = 0.52$. It was selected
due to its red optical and near-infrared colours and borderline $W1-W2$ colour.
This object was determined not to be a quasar, based on the absence of
any broad emission lines or high-redshift spectral features (see Fig\.~\ref{fig:SDSS222934+101907}). 
We classify this target as a K-type star. The position is covered by the FIRST survey,
but no signal is detected from the source as expected for a star.

\subsubsection{SDSS\,J223030+383246 ($z=2.260$)}
This object has colours $g-r = 0.36$, $J-K=1.44$, and $W1-W2 = 1.08$. It was selected
for follow-up spectroscopy to have colours typical of a normal $z<2.8$ quasar.
The redshift of this quasar is determined to be $z = 2.260$. The redshift is measured 
based on the \ion{C}{3}] emission
feature. Although the Ly$\alpha$ and \ion{C}{4} emission lines are also prominent, they
are affected by associated absorption features and therefore are not suitable for Gaussian
fitting. In the SPHEREx spectrum we see a prominent H$\alpha$ line. With the help of 
photometric data, the extinction is determined to be A$_\mathrm{V}$ = 0. We classify this 
target as a normal quasar.

\begin{figure}[htbp]
\centering
\includegraphics[width=0.48\textwidth]{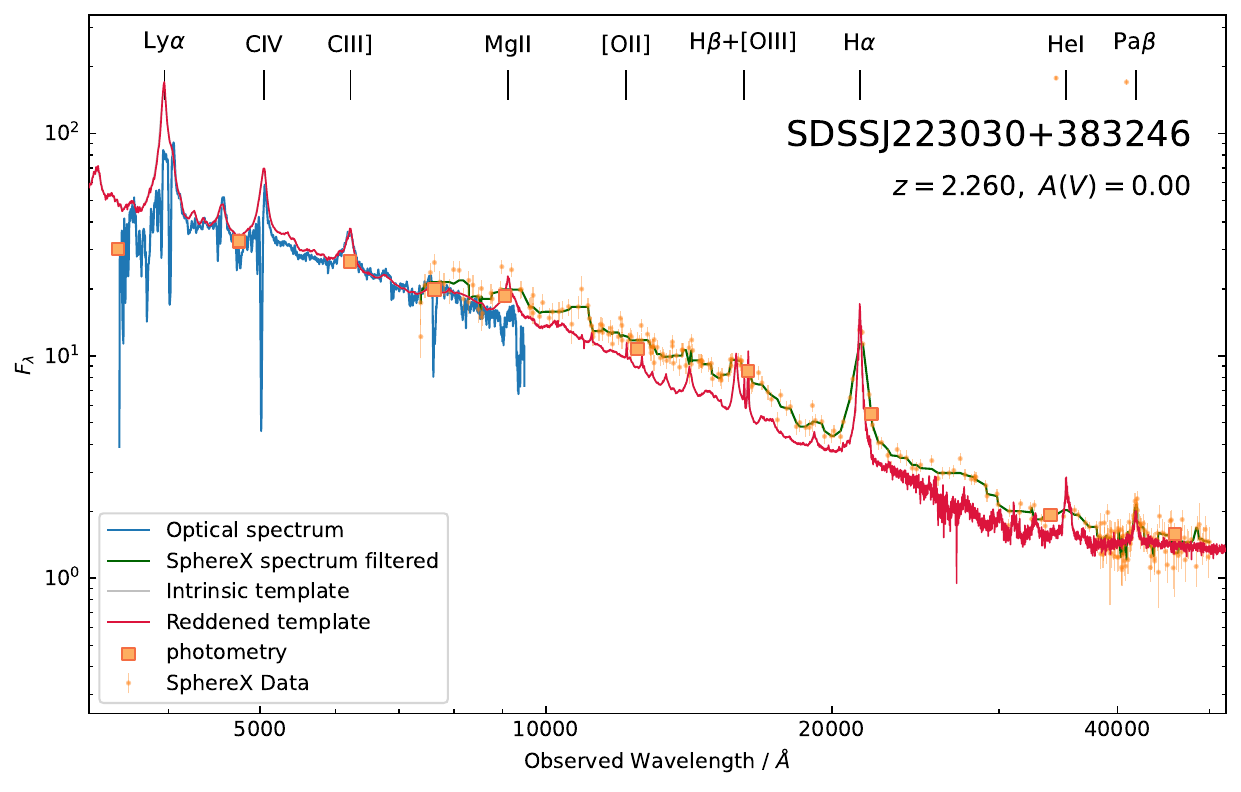}
\caption{The spectrum of SDSS\,J223030+383246. We classify this source as a normal quasar.}
\label{fig:SDSSJ223030+383246}
\end{figure}

\subsubsection{SDSS\,J225432+063826 ($z=2.698$)}

This object has colours
$g-r = 0.92$,
$J-K = 1.24$, and
$W1-W2 = 1.06$.
It was selected for follow-up spectroscopy due to its red colours.
We determine a redshift of 2.698 based on a mix of absorption (Ly$\alpha$
and \ion{C}{4}) and emission (\ion{C}{3}])
features in the optical and emission lines (H$\alpha$ and \ion{He}{1}) in 
the SPHEREx spectrum. The spectrum is mildly reddened. We categorize the 
source as a peculiar quasar due to the weak emission lines in the restframe
ultraviolet. 

\begin{figure}[htbp]
\centering
\includegraphics[width=0.48\textwidth]{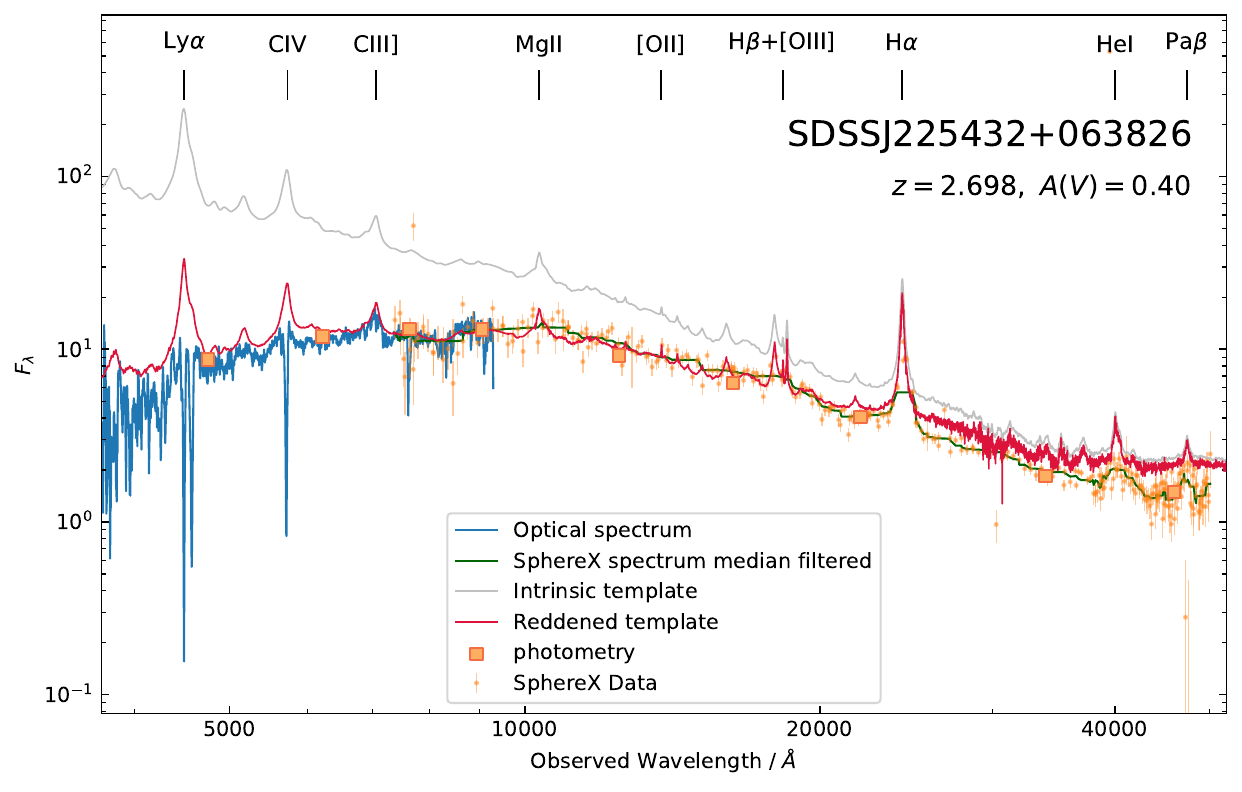}
\caption{The spectrum of SDSS\,J225432+063826. We classify this source as a peculiar quasar.}
\label{fig:SDSSJ225432+063826}
\end{figure}

\subsubsection{SDSS\,J231514+113820 ($z = 2.895$)}
\begin{figure}[htbp]
\centering
\includegraphics[width=0.48\textwidth]{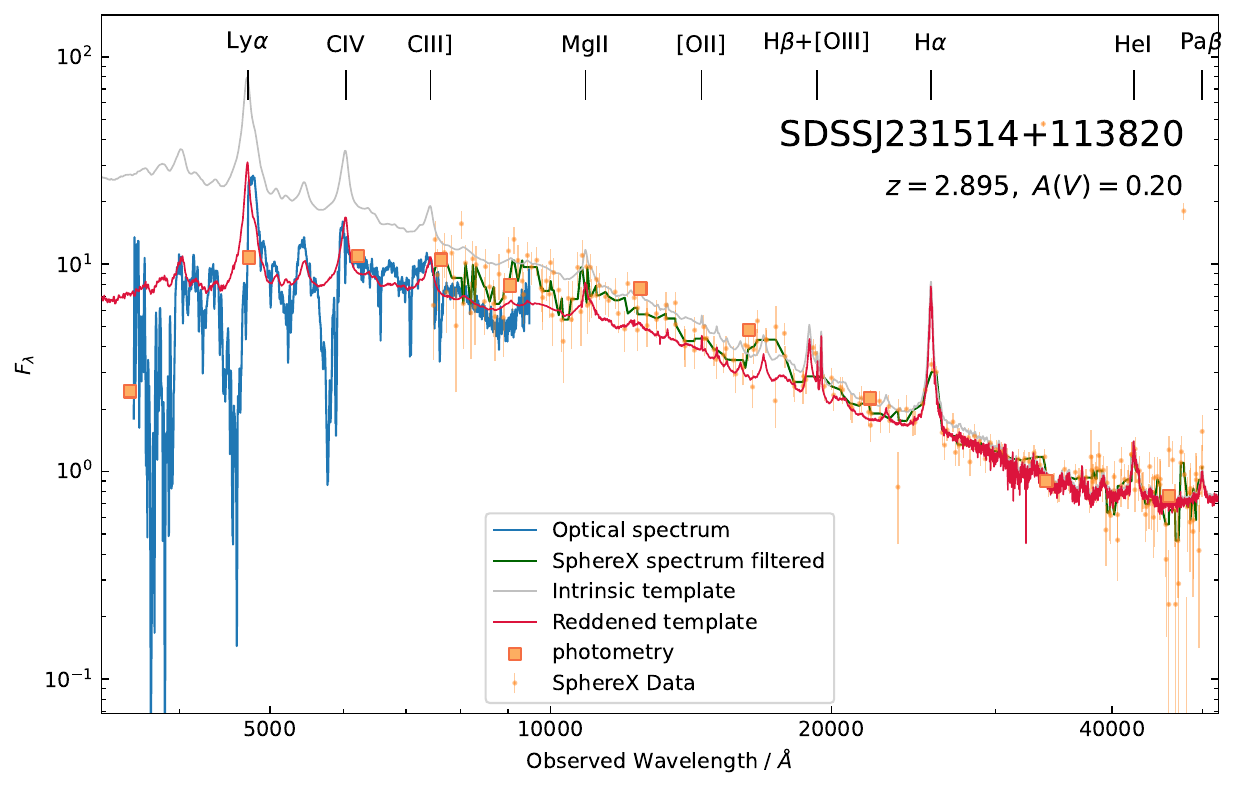}
\caption{The spectrum of SDSS\,J231514+113820. We classify this source as a BAL quasar.}
\label{fig:SDSSJ231514+113820}
\end{figure}

This object has colours $g-r = 0.61$, $J-K=0.85$, and $W1-W2 = 1.11$. It was selected
for follow-up spectroscopy due to its mildly red optical colour.
This  object is a classical BAL quasar (see Fig.~\ref{fig:SDSSJ231514+113820}. Its redshift ($z = 2.895$) 
is inferred from the Ly$\alpha$ and \ion{C}{4} emission features at the red edges of the BAL troughs. 
The SPHEREx spectrum shows emission lines from H$\alpha$ and \ion{He}{1} consistent
with $z=2.895$. The position is covered by the FIRST survey,
but no signal is detected from the source.

\subsubsection{SDSS\,J231704+160807 ($z = 0.464$)}
\begin{figure}[htbp]
\centering
\includegraphics[width=0.48\textwidth]{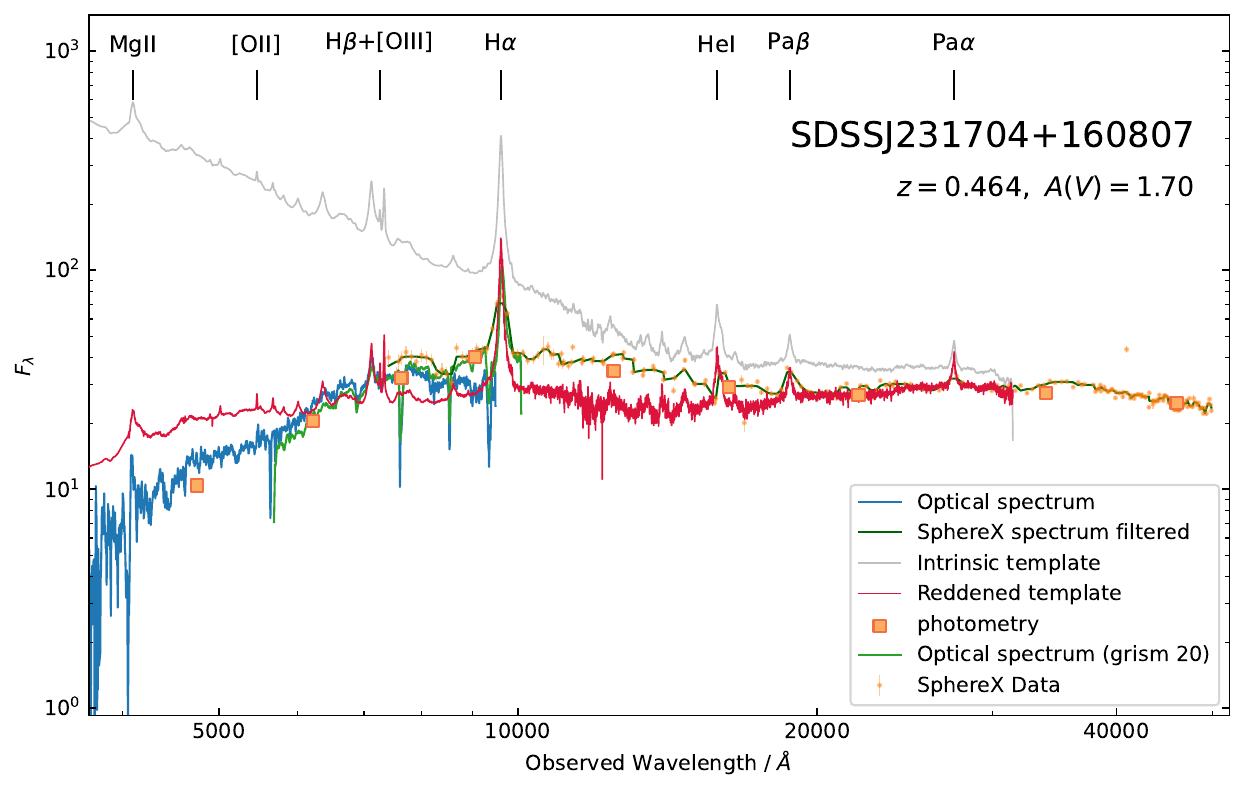}
\caption{The spectrum of SDSS\,J231704+160807. We classify this source as a peculiar quasar.}
\label{fig:SDSSJ231704+160807}
\end{figure}

This object has colours $g-r = 1.32$, $J-K=1.89$ and $W1-W2 = 1.19$. It was selected
for follow-up spectroscopy due to its very red optical colour.
This object is a heavily reddened quasar at a fairly low redshift of $z=0.464$ (see Fig.~\ref{fig:SDSSJ231704+160807}). 
Its redshift is determined based on the H$\alpha$ emission feature round 9600 {\AA}. The
dust extinction is estimated to be A$_\mathrm{V}$ = 1.7, but the match is not very good, especially
in the blue end of the optical range. There seems to be blueshifted absorption (by about 10$^4$ km s$^{-1}$) from both H$\beta$ and H$\alpha$ and possible also less blueshifted absorption from from \ion{He}{1}. Also 
from \ion{Mg}{2} there is blueshifted absorption. In the 
SPHEREx spectrum we see H$\alpha$, \ion{He}{1}, and Paschen lines. We
label the object as a peculiar quasar.

\subsubsection{SDSS\,J231817+234950 ($z=0.570$)}

\begin{figure}[htbp]
\centering
\includegraphics[width=0.48\textwidth]{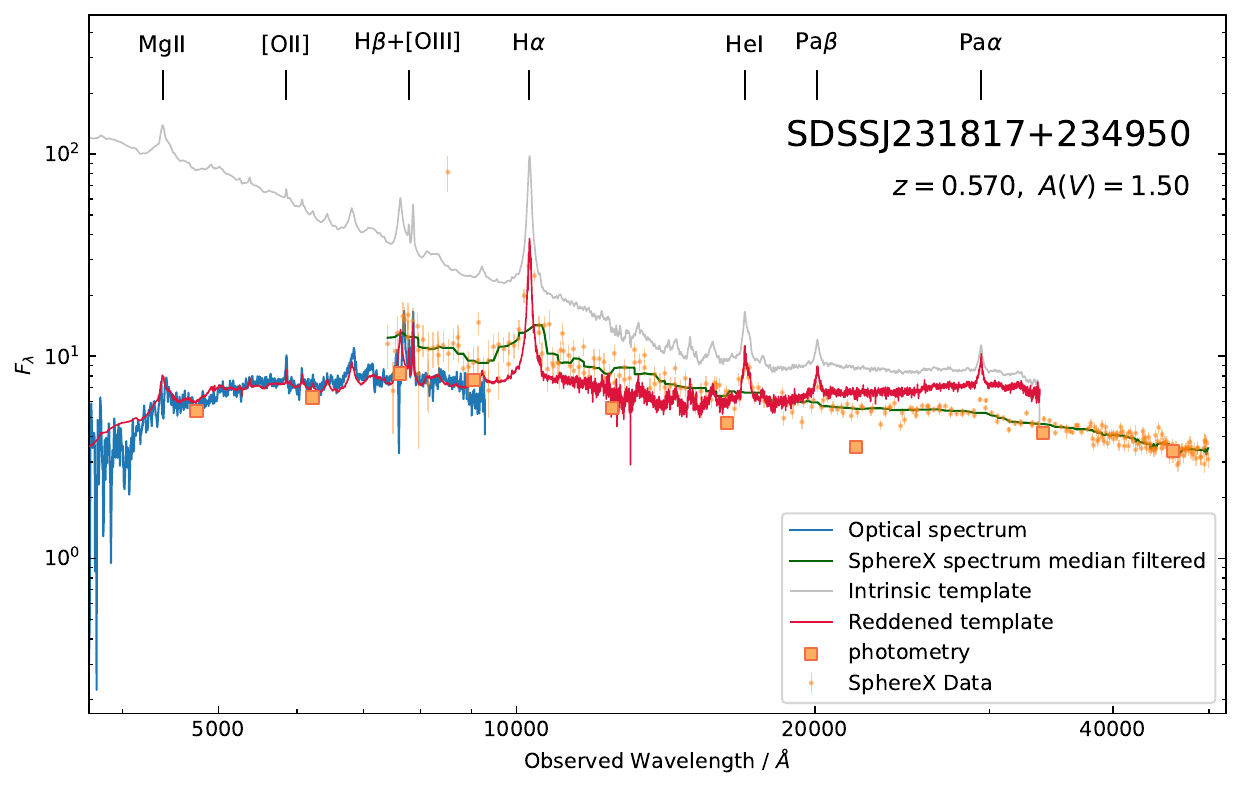}
\caption{The spectrum of SDSS\,J231817+234950. We classify this source as a normal, reddened quasar.}
\label{fig:SDSSJ231817+234950}
\end{figure}

This object has colours
$g-r = 0.76$,
$J-K = 1.69$, and
$W1-W2 = 1.08$.
it was selected for follow-up spectroscopy due to its red
optical colours. We determine a redshift of $z=0.570$ based on a range
of features in the optical and SPHEREx spectra (see Fig.~\ref{fig:SDSSJ231817+234950}. The spectrum is fairly 
well represented by a reddened template with $A_\mathrm{V} = 1.50$. 
We classify this source as a normal, reddened quasar.

\subsubsection{SDSS\,J232521+191557 (star)}
\begin{figure}[htbp]
\centering
\includegraphics[width=0.48\textwidth]{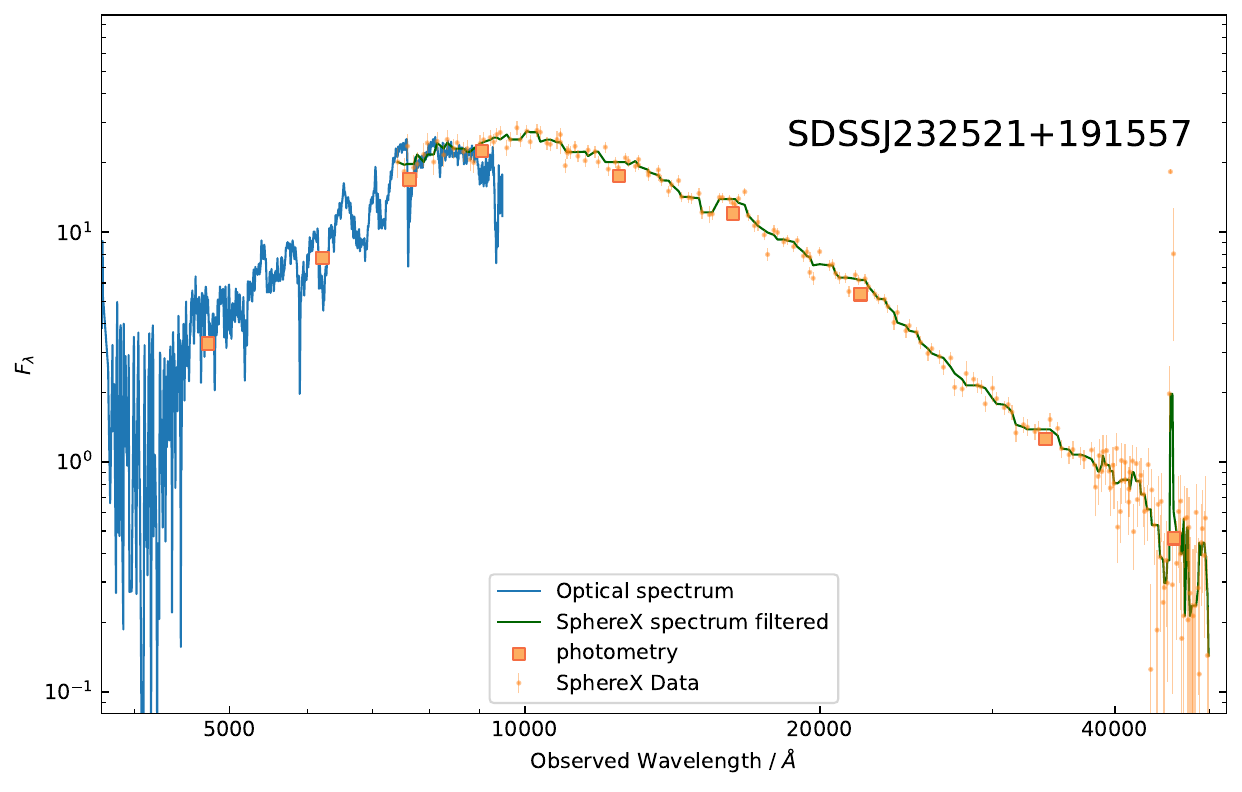}
\caption{The spectrum of SDSS\,J232521+191557. We classify this source as an M-type star.}
\label{fig:SDSSJ232521+191557}
\end{figure}

This object has colours $g-r = 1.51$, $J-K=1.17$, and $W1-W2 = 0.22$. It was selected
due to its red optical colour.
The spectrum of this object lacks the broad emission lines typically observed
in quasars (see Fig.~\ref{fig:SDSSJ232521+191557}). We identify several key features commonly associated with M-type
stars, such as strong TiO absorption bands that shape the overall shape of the
spectrum. We classify this target as an M-type star.

\subsubsection{SDSS\,J233334+291822 ($z=1.675$)}

This object has colours
$g-r = 1.36$,
$J-K = 1.09$, and
$W1-W2 = 0.92$. 
The object was selected for spectroscopy due to its red
optical colours. The spectrum has a single weak emission line 
in the restframe ultraviolet, namely \ion{C}{3}]. There is a BAL line at the position of \ion{C}{4}. In the SPHEREx spectrum we detect H$\alpha$. 
The spectrum is reddened, but the reddened template does not match the 
observed spectrum well bluewards of \ion{Mg}{2}. We classify the object
as a BAL quasar.

\begin{figure}[htbp]
\centering
\includegraphics[width=0.48\textwidth]{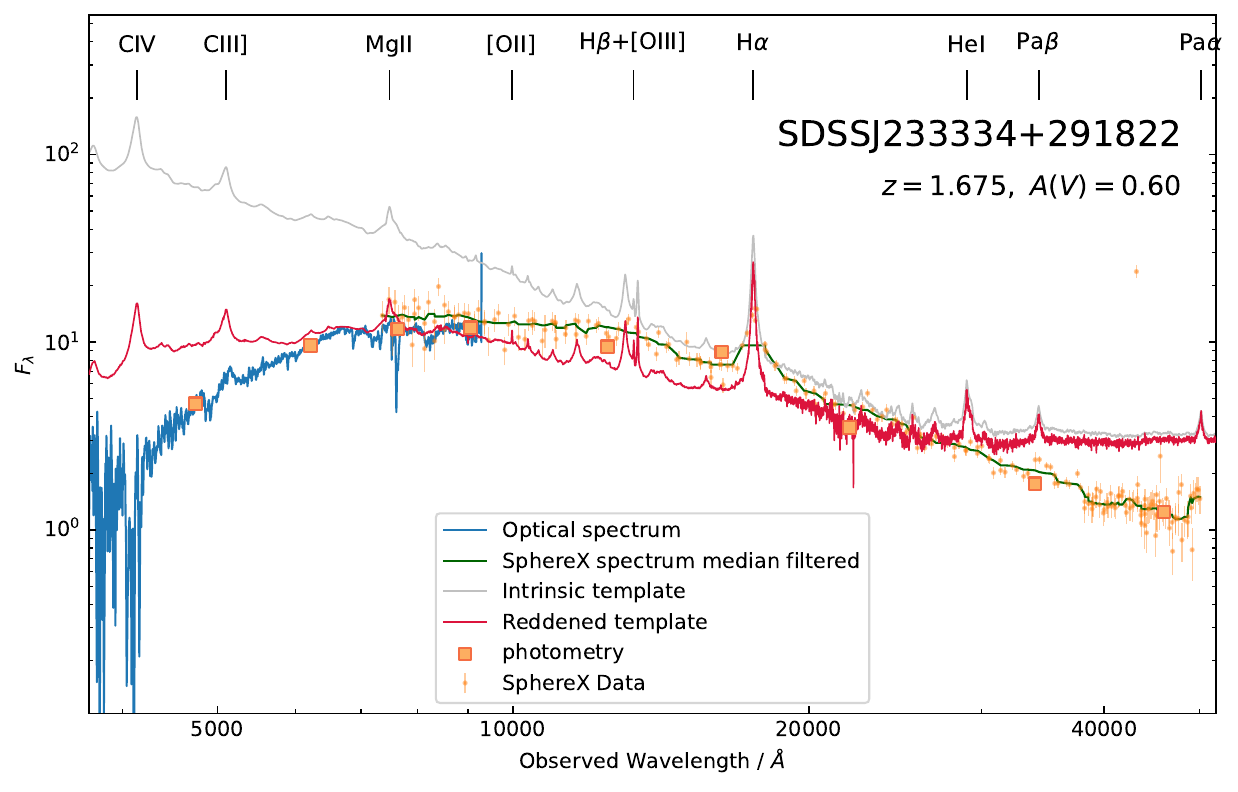}
\caption{The spectrum of SDSS\,J233334+291822. We classify this source as a BAL quasar.}
\label{fig:SDSSJ233334+291822}
\end{figure}

\subsubsection{SDSS\,J234610+431950 ($z = 3.260$)}
\begin{figure}[htbp]
\centering
\includegraphics[width=0.48\textwidth]{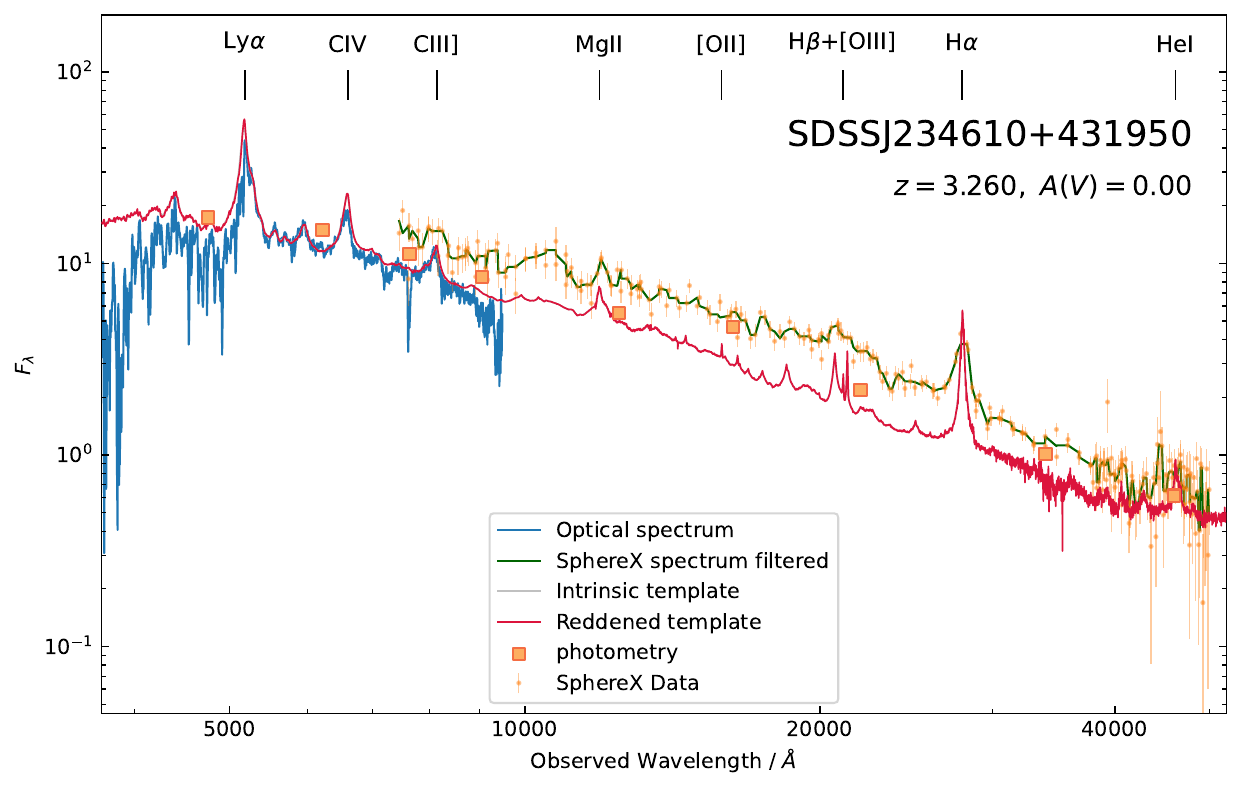}
\caption{The spectrum of SDSS\,J234610+431950. We classify this source as a normal quasar.}
\label{fig:SDSSJ234610+431950}
\end{figure}

This object has colours $g-r = 0.42$ and $W1-W2 = 0.75$. It was selected
for follow-up spectroscopy due to its borderline $W1-W2$ colour suggesting it could be a
quasar at $z>2.8$.
This target is indeed a quasar at a relatively high redshift of $z = 3.260$ (see Fig.~\ref{fig:SDSSJ234610+431950}). 
The redshift is calculated
based on the prominent emission lines from Ly$\alpha$ and \ion{C}{4}. In the SPHEREx spectrum
we also detect a prominent H$\alpha$ emission line. 
We classify it as a normal quasar.

\subsubsection{SDSS\,J234703+373435 ($z = 3.030$)}
\begin{figure}[htbp]
\centering
\includegraphics[width=0.48\textwidth]{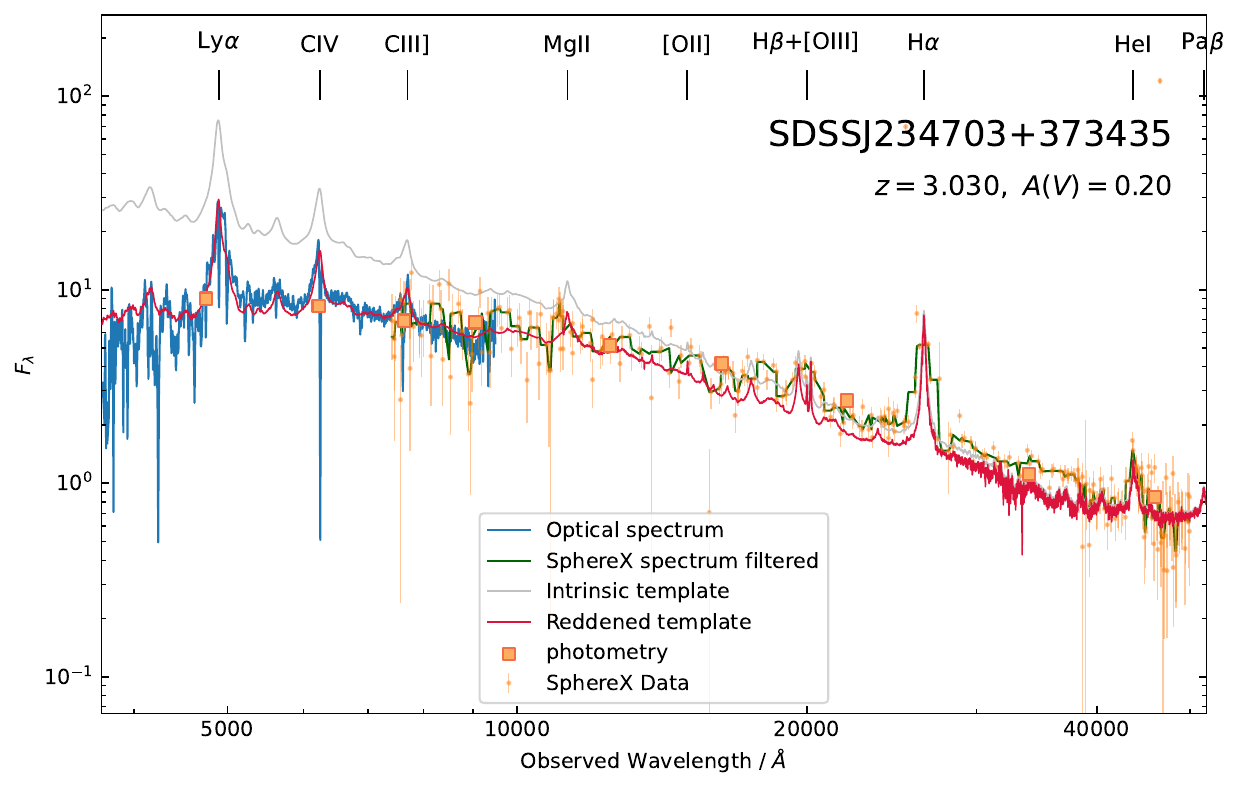}
\caption{The spectrum of SDSS\,J234703+373435. We classify this source as a normal quasar.}
\label{fig:SDSSJ234703+373435}
\end{figure}

This object has colours $g-r = 0.49$, $J-K=1.46$, and $W1-W2 = 1.00$. It was selected
to be a normal quasar.
This object is a quasar with a fairly high redshift $z = 3.030$ (see Fig.~\ref{fig:SDSSJ234703+373435}). 
The redshift is determined based on
the Ly$\alpha$ and \ion{C}{4} emission lines. In the SPHEREx spectrum we detect bright emission lines consistent with H$\alpha$ 
and \ion{He}{1} at this redshift. The dust extinction is estimated to be
A$_\mathrm{V}$ = 0.20. The source is detected by LOFAR with a flux of 1.28$\pm$0.09 mJy.

\subsubsection{SDSS\,J235003+241943 ($z = 0.692$)}

This object has colours 
$g-r = 0.83$,
$J-K = 1.30$, and
$W1-W2 = 1.06$.
The object was selected for follow-up spectroscopy due to its red optical 
colour.
We determine a redshift of $z=0.692$ based on H$\beta$ and [\ion{O}{3}] 
emission the optical and H$\alpha$ in the SPHEREx spectrum. There is 
no line emission from \ion{Mg}{2}. The shape of the 
spectrum is poorly matched by a reddened quasar template. We categorize 
the object as a peculiar quasar. 

\begin{figure}[htbp]
\centering
\includegraphics[width=0.48\textwidth]{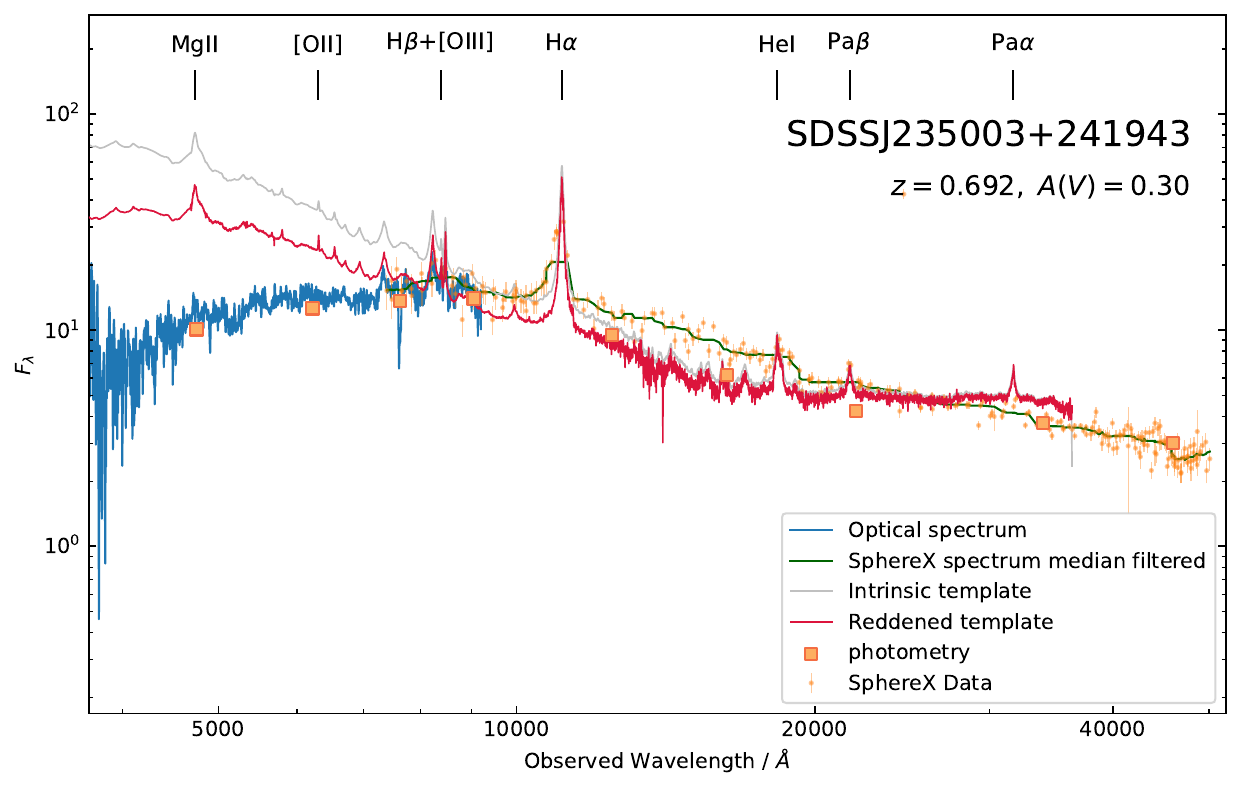}
\caption{The spectrum of SDSS\,J235003+241943. We classify this source as a peculiar quasar.}
\label{fig:SDSSJ235003+241943}
\end{figure}